\def\BEQ{\begin{eqnarray}}
\def\EEQ{\end{eqnarray}}
\def\BML{\begin{mathletters}}
\def\EML{\end{mathletters}}
\def\BE{\begin{equation}}
\def\EE{\end{equation}}

\def\alt{\ {\raise -0.25em \hbox{{$\buildrel < \over \sim $ }}}\ }
\def\agt{\ {\raise -0.25em \hbox{{$\buildrel > \over \sim $ }}}\ }

\documentstyle[prb,aps,epsf]{revtex}
\begin{document}

\input amssym.def
\input amssym

\draft
%\preprint{Rutgers 2000}
\title{The First Temperature Corrections to the Fermi Liquid Fixed Point in
Two Dimensions}
\author{Gennady Y. Chitov and Andrew J. Millis}
\address{Center for Materials Theory}
\address{Department of Physics and Astronomy}
\address{Rutgers University, Piscataway, New Jersey 08854}
\date{\today}
\maketitle
\begin{abstract}
We  calculate using perturbative calculations and Ward identities
the basic parameters of the Fermi Liquid: the scattering
vertex, the Landau interaction function, the effective mass,
specific heat, and physical susceptibilities
for a model of two-dimensional (2D) fermions with a short ranged
interaction at non-zero temperature. 
The leading temperature dependence of the spin components of the
scattering vertex, the Landau function, and the spin
susceptibility is found to be linear. 
We find that the standard $T=0$ relationships for a 
Galilean-invariant Fermi Liquid  are violated by finite-temperature terms.
The coefficients in the temperature corrections
to these relationships involve a subtle interplay between
contributions from small and large ($\sim 2k_F$) momentum processes.
A connection with previous studies
of the 2D Fermi-Liquid parameters is discussed. 
We conclude that
the linear leading temperature dependence of the parameters is a generic
feature of the 2D Fermi Liquid.
\end{abstract}


%
%
%%%%%%%%%%%%%%%%%%%%%%%%%%%%%%%%%%%%%%%%%%%%%%%%%%%%%%%%%%%%%%%%%%%%%%%%%%%%%%
\section{Introduction }\label{Intro}
%%%%%%%%%%%%%%%%%%%%%%%%%%%%%%%%%%%%%%%%%%%%%%%%%%%%%%%%%%%%%%%%%%%%%%%%%%%%%%
%
%
For more than a decade two-dimensional (2D) fermion systems have been a
subject of great interest. One important motivation has been the
non-Fermi-Liquid behavior observed in high-$T_c$ superconductors
above $T_c$.\cite{HighTc}
In this context the existence and stability of the Fermi Liquid  (FL) 
in $d > 1$ has been extensively investigated within various 
modern approaches.\cite{Shankar94,GD95,Houghton,Metzner,Mat} 
It seems fair to say that no compelling theoretical evidence has been found
for non-Fermi-Liquid behavior in systems with short ranged
interactions and in the absence of coupling to gauge fields. 
The FL is stable, provided that standard conditions\cite{AGD,Lifshitz80} 
are satisfied.

Given this a natural question arises: suppose 2D fermions are in  the FL 
phase in the sense of the Landau's Fermi Liquid Theory
(FLT).\cite{Landau,AGD,Lifshitz80}
We understand ``zero order'' FL quantities  as those defined in the limit of 
zero temperature and zero momentum-energy transfers. Then what is the 
low-energy  behavior of such quantities? In this paper we study the leading 
temperature corrections to the zero-order behavior of the the components 
of the Landau function and the scattering vertex, the quasiparticle's effective 
mass, and the uniform response functions (compressibility, spin susceptibility).
In Renormalization Group (RG) language  we are studying  the  corrections 
coming from the leading irrelevant terms.\cite{Shankar94,GD95,GD98}

Interest in  the issue grown in recent years also because these
leading corrections provide the ``bare'' temperature dependence
of the parameters in theories describing quantum critical
phenomena in metals.\cite{HeM} In particular, puzzling data in
several materials \cite{CeCu6,Lonzarich} led to explanations
based on an unusual underlying temperature, momentum or frequency
dependence of electronic susceptibilities.\cite{Coleman98,Si00}

The leading temperature corrections to the parameters of a {\it stable} FL
been studied for many years. Rather surprisingly,
the issue  remains a subject of controversy.
For example, it was found\cite{DonAmit67}
that the leading temperature correction to the specific heat
coefficient $\gamma=C/T$ was $T^{2}\ln T$ in d=3 spatial
dimensions and $T$ in d=2.\cite{Bedell93,Misawa99}.
Recently these results were also rederived by multidimensional bosonization
(see, e.g., Ref.~[\onlinecite{Houghton}]).
Whether the spin and charge susceptibilities display
similarly anomalous (i.e., non-$T^2$) temperature dependence
is the  subject of a lengthy controversy in the literature:
see, e.g., Ref.~[\onlinecite{CP77}], discussion and references therein.
For a most recent reassesment of such results see Ref.~[\onlinecite{BK97}].
The prevailing conclusion  was that of
Carneiro and Pethick\cite{CP77} who found no leading
$T^2 \ln T$ correction to the spin susceptibility of the 3D FL.
Their analysis implies that terms $\propto T$ are absent in 2D.

The heuristic argument which runs commonly through the literature
to account {\it a posteriori} for the absence of anomalous terms 
(in $T$ or in $q$) in response functions is that although these terms 
are known to occur in the individual self-energy and vertex diagrams, 
they cancel in final results for symmetry reasons mathematically expressed 
via Ward identities. As we now discuss, this argument is  misleading.
Although exact symmetries 
of a model (e.g., global gauge and/or rotational invariance)
result in relationships between the self-energy and the
vertex through the Ward identities (cf. Sec.~\ref{WIsec} below),
these are  usually not enough to close the system of equations and
demonstrate explicitly the cancellation of the self-energy and
vertex corrections. Additional symmetries like, e.g., the chiral symmetry of 
the 1D Tomonaga-Luttinger model, are needed  in order to do it.\cite{DL74} 
In $d>1$ analogous extra {\it asymptotic} symmetries first noted by 
Haldane\cite{Haldane} occur in the fermionic Wilsonian 
{\it low-energy effective action} under some model's restrictions.
In the recent series of papers by Metzner, Castellani and Di Castro 
(reviewed in Ref.~[\onlinecite{Metzner}]) the relationship between the 
extra asymptotic symmetry of the effective action 
and the RPA (FLT) results for the response functions was clearly 
established. They also clarified a close connection between this approach 
and multidimensional bosonization.\cite{Houghton}
The additional Ward identities are properties of a model in which 
all processes involving momentum transfers greater than a cutoff
\begin{equation}
\label{Lamea}
  \Lambda_{\rm ea} \ll k_F
\end{equation}
are discarded.
These restrictions appear naturally in the Wilsonian low-energy effective 
action with the UV cutoff (\ref{Lamea}) obtained from an RG approach such as 
that of Shankar,\cite{Shankar94} in which interacting fermions are treated 
by progressive elimination of modes towards the Fermi surface.
Only asymptotically (i.e., in the limit  $\Lambda_{\rm ea} / k_F \to 0$, 
$T=0$) this forward-scattering action  possesses extra symmetries 
$U(1)^{\infty}$ (and $SU(2)^{\infty}$ for rotationally invariant 
case).\cite{Haldane,Houghton} 
(It is assumed that the initial {\it microscopic} action (Hamiltonian) has
only ordinary $U(1)$ [and $SU(2)$] symmetry.)
However, these extra identities do not constrain
possible $T$- or $q$- terms in physical quantities, coming from irrelevant 
terms. Even in 1D, e.g.,  the $SU(2)_L \otimes SU(2)_R$-breaking marginally
irrelevant term coming from backscattering results in  $1/\ln T$-leading
correction to the uniform spin susceptibility of the Luttinger 
liquid.\cite{1DLut} In higher dimensions there are always 
[even in the limit (\ref{Lamea})] special configurations of momenta 
near the Fermi surface which give rise to irrelevant terms in the 
effective action which strongly interfere with the forward 
scattering channel\cite{GD98} and invalidate asymptotic symmetries. 
In the framework of the low-energy effective action approach there is 
also another potential source of temperature corrections, i.e., 
the possibility of $T$-dependence of the Wilsonian action's vertices
(couplings) developing on the previous stages of modes elimination
before one  reaches the effective action scale (\ref{Lamea}). Apparently, 
this question has not been carefully studied.

We note that in the most of existing literature studying the leading 
FL corrections it is assumed that the crucial coupling is  between
quasiparticles and long-wavelength collective modes, i.e., only small
momentum scales are taken into account.
However the possibility of ``$2k_{F}$ singularities'',
i.e., anomalous temperature terms coming from processes involving large
($\sim 2k_{F}$) momentum transfers, has been pointed out by Misawa for 3D FL 
already in early 70s.\cite{Misawa3D} Apparently due to the lack of 
experimental evidence of a $T^2 \ln T$ term in the susceptibility 
of a generic 3D FL and also because Misawa's results rely on
the analysis of selected diagrams [cf. previous 
paragraph], they were widely disregarded  in favor of those of 
Carneiro and Pethick.
In the context of semiconductor physics Stern was the first to 
note\cite{Stern80} that in a 2D electron gas the electron scattering rate 
was proportional to $T$ due to $2k_{F}$ effects.  The consequences of 
the $2k_{F}$ effects for the leading $T$-dependence of 2D FL quantities
have not been considered in the literature until recently.

The issue of the leading correction to FLT has recently been revived by 
several papers. 
Belitz, Kirkpatrick and Vojta\cite{BK97} presented perturbative
calculations, mode-coupling arguments and power counting estimates
which showed that the leading $q$ dependence of the spin susceptibility (but
not the charge susceptibility) was $|q|$ in 2D  ($q^{2}\ln q$ in 3D).
They did not find the analogous $T$-correction explicitly, but 
concluded that one should generally expect a linear $T$-term 
in the 2D FL susceptibility ($T^2\ln T$ in 3D). This dependence has 
important implications  for the theory of the quantum
critical metallic ferromagnet.\cite{BeMi} 

S{\'e}n{\'e}chal and one of us\cite{GD98} predicted 
the occurence of the linear $T$-corrections to the FL vertices from 
one-loop RG  calculations based on a 2D effective action.
Those RG calculations of the irrelevant corrections rely more 
on the low-energy effective action's phase space constrains in their way 
to sort out the effective interactions and the scales involved, rather 
than on doing it according to the strength of the latter, as a perturbation 
theory does. That makes desirable to easily obtain a perturbative signal 
of the RG predictions. Also, an analogous RG calculation of 
other FL quantities was not done.

Hirashima and Takahashi\cite{HT98} performed numerical analyses of 
perturbative expressions which appeared to confirm the 
prediction of Belitz {\it et al}\cite{BK97} for 2D susceptibility.
However due to numerical difficulties in handling divergences 
in some terms they were  unable even to determine the sign of the 
coefficient in the leading $q$-term. Also contrary to Belitz {\it et al} 
who focused on the long-wavelength contributions, 
the authors of Ref.~[\onlinecite{BK97}] emphasized the crucial role of 
$2k_{F}$ contributions  in their findings. Following this Misawa conjectured 
a phenomenological form for the free energy\cite{Misawa99} which 
results in the linear $T$-term in the 2D spin susceptibility and in the 
coefficient $\gamma$, and agrees 
well with the numerical calculations in lowest order.\cite{HK99}

This previous work has left a number of important questions 
unresolved, including analytic treatment of anomalous terms
(important for verifying cancellations between different
contributions), the relationship to FLT and to Ward identities, 
and the connection to the extensive literature from the 1960-70s.
To elucidate the issues in the most transparent way, we apply the
perturbation theory for 2D contact-interacting spin-$\frac12$
fermions, starting from a {\it microscopic} action. Although
the Landau  FLT is not a perturbative theory,
for sufficiently weak interactions (assuming the
interaction being repulsive and we are above the Kohn-Luttinger
temperature) one should be able to find  the parameters of the
FL in terms of the coupling series. 

We present what is apparently the first analytic calculation of the leading 
$T$-dependence of the effective mass, Landau parameters and response functions
of a 2D electron gas, to second order in the interaction
strength, {\it including all channels and all momentum processes (scales)}.
We take into account the Ward identities explicitly.
As we show in this paper, the processes involving large ($\sim
2k_{F}$) momentum transfers are crucial to the anomalous temperature dependence
of FL quantities.

The rest of this paper is organized as follows. Section \ref
{Model} defines the model. Section \ref{Vertex} defines the four-point
vertex to be calculated and its relationship to the Fermi Liquid parameters.
We also give there the basic equation for that vertex in the one-loop
approximation. In Section \ref{ScatLan} we present and discuss our results
for the scattering amplitude and the Landau function. In Section \ref{WIsec}
we give  the Ward identities which are used for the following calculations.
In Section \ref{efmass} we calculate the effective mass. Sections \ref{compr}, 
\ref{susc} present the results for the compressibility and the spin
susceptibility, respectively. The results are recapitulated in the
concluding Section \ref{Concl}. Appendices A and B contain detailed
presentation of the calculations of the one-loop contributions entering the
equation for the vertex. One of the goals of presenting these technicalities
was to give the idea of how more involved two- and three-loop calculations
of Secs. \ref{efmass}-\ref{susc} were carried out.

Some of the results presented here were announced in a previous short
communication.\cite{GAlet}
%
%
%%%%%%%%%%%%%%%%%%%%%%%%%%%%%%%%%%%%%%%%%%%%%%%%%%%%%%%%%%%%%%%%%%%%%%%%%%%%%%
\section{Model and Formalism}\label{MaF}
%%%%%%%%%%%%%%%%%%%%%%%%%%%%%%%%%%%%%%%%%%%%%%%%%%%%%%%%%%%%%%%%%%%%%%%%%%%%%%
%
%
\subsection{Model}\label{Model}
%%%%%%%%%%%%%%%%%%%%%%%%%%%%%%%%%%%%%%%%%%%%%%%%%%%%%%%%%%%%%%%%%%%%%%%%%%%%%%
We treat interacting fermions at finite temperature in the
standard path integral Grassmannian formalism.\cite{Negele88}
The partition function is
\begin{equation}
\label{Z}
{\cal Z} =\int {\cal D}\bar\psi{\cal D}\psi~e^{S_0 + S_{\rm int}}
\end{equation}
where the free part of the action is
\begin{equation}
\label{S0}
S_0 = \int_{({\bf 1})}\bar\psi_{\alpha}({\bf 1})\left[
i\omega_1+\mu-\epsilon({\bf k}_1)\right]\psi_{\alpha}({\bf 1})
\end{equation}
and 
\begin{mathletters}
\label{notation}
\begin{eqnarray}
\int_{({\bf i})}~~&\equiv&~~{1\over\beta}\int{d{\bf k}_i\over(2\pi)^d}
\sum_{\omega_i}\\
({\bf i}) ~~&\equiv&~~ ({\bf k}_i,\omega_i) ~~\equiv~~ {\cal K}_i
\end{eqnarray}
\end{mathletters}
where $\beta$ is the inverse temperature, $\mu$ the chemical potential,
$\omega_i$ the fermion Matsubara frequencies and $\psi_{\alpha}({\bf i})$ a
$N$-component Grassmann field with a spin (flavor, if $N \neq 2$) index
$\alpha$. Summation over repeated indices is implicit throughout this
paper. We set $k_B=1$ and $\hbar=1$.
The $SU(N)$-invariant quartic interaction is 
\begin{equation}
\label{Sint}
S_{\rm int} = -\frac14\int_{({\bf 1},{\bf 2},{\bf 3},{\bf 4})}
\bar\psi_{\alpha}({\bf 1})\bar\psi_{\beta}({\bf 2})
\psi_{\gamma}({\bf 3})\psi_{\varepsilon}({\bf 4})
U^{\alpha\beta}_{\gamma\varepsilon}({\bf 1},{\bf 2};{\bf 3},{\bf 4})
\delta^{(d+1)}({\bf 1}+{\bf 2}-{\bf 3}-{\bf 4})
\end{equation}
Here the conservation of energy and momentum is enforced by the symbolic delta
function
\begin{equation}
\label{Delta}
\delta^{(d+1)}({\bf 1}+{\bf 2}-{\bf 3}-{\bf 4})\equiv\beta(2\pi)^d
\delta({\bf k}_1+{\bf k}_2-{\bf k}_3-{\bf k}_4)
\delta_{\omega_1+\omega_2-\omega_3-\omega_4,0}
\end{equation}
The $SU(N)$ extension of the physical $SU(2)$ symmetry  is useful for
different applications,\cite{GD95} and the case of the spin-$\frac12$
electrons is recovered by setting $N=2$.
We can decompose  the interaction by factorizing  its symmetric and antisymmetric
frequency-momentum- and spin- dependent  parts in the way which
explicitly manifests the antisymmetry of the function $U$ under
exchange:\cite{GD95}
\begin{equation}
\label{potS}
U^{\alpha\beta}_{\gamma\varepsilon} = U^A I^{\alpha\beta}_{\gamma\varepsilon}
+ U^S T^{\alpha\beta}_{\gamma\varepsilon}
\end{equation}
where the functions $U^S$ and $U^A$ have the symmetry properties
\begin{mathletters}
\label{USYM}
\begin{eqnarray}
U^A({\bf 1},{\bf 2};{\bf 3},{\bf 4}) &=&
-U^A({\bf 2},{\bf 1};{\bf 3},{\bf 4}) =
-U^A({\bf 1},{\bf 2};{\bf 4},{\bf 3})\\
U^S({\bf 1},{\bf 2};{\bf 3},{\bf 4}) &=&
\phantom{-} U^S({\bf 2},{\bf 1};{\bf 3},{\bf 4}) =
\phantom{-} U^S({\bf 1},{\bf 2};{\bf 4},{\bf 3})
\end{eqnarray}
\end{mathletters}
while two operators $\hat I$ and $\hat T$, which are  respectively symmetric and
antisymmetric in the spin  space, are defined as follows:
\begin{mathletters}
\label{base}
\begin{eqnarray}
I^{\alpha\beta}_{\gamma\varepsilon}~&\equiv&~
\delta_{\alpha\varepsilon}\delta_{\beta\gamma} +
\delta_{\alpha\gamma}\delta_{\beta\varepsilon}\\
T^{\alpha\beta}_{\gamma\varepsilon}~&\equiv&~
\delta_{\alpha\varepsilon}\delta_{\beta\gamma} -
\delta_{\alpha\gamma}\delta_{\beta\varepsilon}
\end{eqnarray}
\end{mathletters}

If we assume an instantaneous point interaction between electrons, then
$U^A=0$ and $U^S=U_0$. The sign of the interaction is chosen such that
$U_0>0$ corresponds to repulsion. Then
\begin{equation}
\label{SintHub}
S_{\rm int}^c =- \frac14 U_0 \int_{({\bf 1},{\bf 2},{\bf 3},{\bf 4})}
\bar\psi_{\alpha}({\bf 1})\bar\psi_{\beta}({\bf 2})
\psi_{\gamma}({\bf 3})\psi_{\varepsilon}({\bf 4})
T^{\alpha\beta}_{\gamma\varepsilon}
\delta^{(d+1)}({\bf 1}+{\bf 2}-{\bf 3}-{\bf 4})
\end{equation}
In this paper we consider mainly 2D electrons with the bare spectrum of a 
free gas $\epsilon({\bf k})=k^2/2m$ and  the circular Fermi surface,
interacting via (weak) contact repulsion (\ref{SintHub}), 
but discuss the consequences of generic spectra and a
non-circular Fermi surface.
%
%
%%%%%%%%%%%%%%%%%%%%%%%%%%%%%%%%%%%%%%%%%%%%%%%%%%%%%%%%%%%%%%%%%%%%%%%%%%%%%%
\subsection{Four-point 1PI vertex and parameters of Fermi Liquid}\label{Vertex}
%%%%%%%%%%%%%%%%%%%%%%%%%%%%%%%%%%%%%%%%%%%%%%%%%%%%%%%%%%%%%%%%%%%%%%%%%%%%%%
%
%
We define the two-particle Green's function as:
\begin{equation}
\label{G2}
{\hat G}_2({\bf 1},{\bf 2};{\bf 3},{\bf 4}) = -
\langle\psi_{\alpha}({\bf 1})\psi_{\beta}({\bf 2})
\bar\psi_{\gamma}({\bf 3})\bar\psi_{\varepsilon}({\bf4})\rangle
\end{equation}
The 1PI vertex  ${\hat\Gamma}$ is related to ${\hat G}_2$ as
\begin{eqnarray}
\label{2PV}
&~& {\hat G}_2({\bf 1},{\bf 2};{\bf 3},{\bf 4}) =
\langle \psi_{\alpha}({\bf 1})  \bar \psi_{\gamma}({\bf 3}) \rangle
\langle \psi_{\beta}({\bf 2}) \bar \psi_{\varepsilon}({\bf 4}) \rangle
-
\langle \psi_{\alpha}({\bf 1}) \bar \psi_{\varepsilon}({\bf 4}) \rangle
\langle \psi_{\beta}({\bf 2})  \bar \psi_{\gamma}({\bf 3}) \rangle
\nonumber \\
&+& \int_{{\bf 1}',{\bf 2}'{\bf 3}'{\bf 4}'}
\langle  \psi_{\alpha}({\bf 1}) \bar \psi_{\alpha '}({\bf 1 '}) \rangle
\langle \psi_{\beta}({\bf 2}) \bar   \psi_{\beta '}({\bf 2 '}) \rangle
\Gamma^{\alpha ' \beta '}_{\gamma ' \varepsilon '}
({\bf 1 '},{\bf 2 '};{\bf 3 '},{\bf 4 '})
\langle \psi_{\gamma '}({\bf 3 '}) \bar \psi_{\gamma}({\bf 3}) \rangle
\langle \psi_{\varepsilon '}({\bf 4 '})
\bar \psi_{\varepsilon}({\bf 4}) \rangle
\nonumber \\
~&~&~~~~~~~~~~~~\times\delta^{(d+1)}({\bf 1 '}+{\bf 2 '}-{\bf 3 '}-{\bf 4 '})
\end{eqnarray}
where due to the symmetries of the model we can write
the one-particle Green's function as
\begin{equation}
\label{GFF}
- \langle  \psi_{\alpha}({\bf 1}) \bar \psi_{\beta}({\bf 2}) \rangle
=G({\bf 1})\delta_{\alpha \beta}
\delta^{(d+1)}({\bf 1 }-{\bf 2 })
\end{equation}
Perturbatively, the vertex is constructed by considering only connected
one-particle-irreducible (1PI)
diagrams with amputated external legs. In lowest order
${\hat\Gamma}^{(0)}({\bf 1},{\bf 2};{\bf 3},{\bf 4}) =
{\hat U}({\bf 1},{\bf 2};{\bf 3},{\bf 4})$. To shorten notations we use
the hat meaning that it is an operator in the spin space,
and it also comprises two components.
Along with the representation (\ref{potS})
with the components ($\Gamma^A$, $\Gamma^S$)
for the vertex, which explicitly shows its antisymmetry, we will also
use another representation more convenient for some applications, 
namely separating the vertex into charge and spin components via
\begin{equation}
\label{Vrep}
\hat \Gamma \mapsto
\Gamma^{\alpha \beta}_{\gamma \varepsilon}=
\Gamma^A I^{\alpha \beta}_{\gamma \varepsilon}+
\Gamma^S T^{\alpha \beta}_{\gamma \varepsilon}=
-{1 \over N} \Gamma_{\rm ch}\delta_{\alpha \gamma}\delta_{\beta \varepsilon}
-\frac12 \Gamma_{\rm sp} 
\lambda^a_{\alpha \gamma}\lambda^a_{\beta \varepsilon}
\end{equation}
where
$\hat \lambda^a$ ($a=1,...,N^2-1$) are Hermitian traceless generators of the
$SU(N)$ group, coinciding with the Pauli matrices for the case $N=2$, and
normalized such that
\begin{equation}
\label{LamNor}
{\rm tr}(\hat \lambda^a\hat \lambda^b)=2\delta^{ab}
\end{equation}
The components of the vertex in different representations are related as
\begin{mathletters}
\label{ABdef}
\begin{eqnarray}
\Gamma_{\rm ch} &=& (N-1) \Gamma^S-(N+1) \Gamma^A \\
\Gamma_{\rm sp}  &=& - \Gamma^S- \Gamma^A
\end{eqnarray}
\end{mathletters}
Taking into account momentum and energy conservation, we write the
vertex as:
\begin{equation}
\label{GammaTr}
\hat \Gamma({\bf 1},{\bf 2};{\bf 1}+{\cal Q},{\bf 2}-{\cal Q})
\equiv \hat \Gamma({\bf 1},{\bf 2};{\cal Q}) ~,
\end{equation}
with the  transfer vector
\begin{equation}
\label{Transfer}
{\cal Q} = {\bf 3}-{\bf 1}  \equiv ({\bf q},i\Omega)
\end{equation}
where $i\Omega $ is a bosonic Matsubara frequency.
In what follows we will frequently use the property of the components
$\Gamma_{\rm ch}$ and $\Gamma_{\rm sp}$ which is a consequence of 
the vertex's antisymmetry
\begin{equation}
\label{versym}
\Gamma^{\alpha \beta}_{\gamma \varepsilon}({\bf 1},{\bf 2};{\cal Q})=
\Gamma^{\beta \alpha}_{\varepsilon \gamma}({\bf 2},{\bf 1};{-\cal Q})
~~\Longleftrightarrow~~\Gamma_{\rm ch}({\bf 1},{\bf 2};{\cal Q})=
\Gamma_{\rm ch}({\bf 2},{\bf 1};-{\cal Q}),~~
\Gamma_{\rm sp}({\bf 1},{\bf 2};{\cal Q})=
\Gamma_{\rm sp}({\bf 2},{\bf 1};-{\cal Q})
\end{equation}
For the calculation of physical quantities of the FL we need 
$\hat \Gamma({\bf 1},{\bf 2};{\cal Q})$ at
${\cal Q}=0$. As is well known\cite{Landau,AGD,Lifshitz80}
the limit ${\cal Q} \to 0$ is not unique, since the vertex is
non-analytic function at  ${\cal Q}= 0$.
Two  vertices
\begin{equation}
\label{two-limits1}
\hat \Gamma^q({\bf 1},{\bf 2}) = \lim_{q \to 0}\Bigl
\lbrack \hat \Gamma({\bf 1},{\bf 2};{\cal Q})
\Bigl \vert_{\Omega =0} \Bigr \rbrack \,,
\end{equation}
\begin{equation}
\label{two-limits2}
\hat \Gamma^\Omega({\bf 1},{\bf 2}) = \lim_{\Omega\to0}
\Bigl\lbrack \hat \Gamma({\bf 1},{\bf 2};{\cal Q}) \Bigl
\vert_{q=0} \Bigr \rbrack ~.
\end{equation}
can be  defined unambiguously at ${\bf 1} \neq {\bf 2}$ and then 
continued to ${\bf 1} \to {\bf 2}$. (See, e.g., 
Refs.~[\onlinecite{Lifshitz80,GD98}] on this subtlety).
In the calculation of the FLT vertices (real electrons, $N=2$)
the external momenta are chosen to lie on the Fermi surface, and
the external Fermionic frequencies are put equal to the minimal
Matsubara frequency $i \pi T$. In the case of a rotationally
invariant Fermi surface the vertex dependence on the external
momenta ${\bf k}_1$ and ${\bf k}_2$ lying on the Fermi surface
can be parameterized by the relative angle
$\theta_{12}$ between those momenta and then 
$\Gamma^q(\theta_{12})$, $\Gamma^\Omega(\theta_{12})$ can be
identified with the scattering vertex and the Landau function,
respectively. Namely, for the components of the scattering vertex
($A,~B)$ we have (cf., e.g., Ref.[\onlinecite{Lifshitz80}])
\begin{equation}
\label{scat-comp}
-2 \nu_R Z^2  \Gamma^{\alpha \beta (q)}_{\gamma \varepsilon}(\theta_{12})
=A(\theta_{12}) \delta_{\alpha \gamma }\delta_{\beta \varepsilon }
+B(\theta_{12}) \lambda_{\alpha \gamma }^a \lambda_{\beta \varepsilon }^a
\end{equation}
where $\nu_R=S_d K_F^{d-2}m^{\ast}/(2\pi)^d$ is the
$d$-dimensional renormalized electron density of states per spin on the
Fermi level, $S_d$ is the $d$-dimensional area of the unit sphere
(for $d=2$:  $\nu_R=m^{\ast}/2\pi$), and
$Z$ is the field renormalization constant. So
\begin{equation}
\label{A}
A(\theta_{12})= \nu_R Z^2 \Gamma_{\rm ch}^q(\theta_{12})
\end{equation}
\begin{equation}
\label{B}
B(\theta_{12})= \nu_R  Z^2 \Gamma_{\rm sp}^q(\theta_{12})
\end{equation}
Two components ($F,~G)$ of the Landau interaction function are defined by
analogous equations, with the substitution
$q \mapsto \Omega$, $A \mapsto F$, $B \mapsto G$.

The one-loop approximation $\hat\Gamma^{(1)}$ of the straightforward perturbation 
theory for ${\hat\Gamma}({\bf 1},{\bf 2};{\cal Q})$
in diagrammatic form is given in Fig.~1. For the model given by 
(\ref{S0},\ref{SintHub}) we  obtain after some spin summations
\begin{equation}
\label{GammaPertExp}
\hat\Gamma^{(1)}({\bf 1},{\bf 2};{\cal Q})=
-\frac12 L_{-} U_0^2 \hat I+
\bigg\{
U_0 -\Big[ \frac12 L_{+}+ {\Bbb C}({\bf 1}+{\bf 2}) \Big] U_0^2
\bigg\} \hat T
\end{equation}
with
\begin{equation}
\label{chipm}
L_{\pm} \equiv L({\cal Q}) \pm L({\cal Q}')
\end{equation}
We denote ${\cal Q'} \equiv {\bf 2}-{\bf 1}-{\cal Q}$. The functions
$L$ and ${\Bbb C}$ coming from  the contributions of bubbles (ZS, ZS') and
(BCS), respectively, are defined and calculated in Appendices \ref{ApA}
and \ref{ApB}.
From the one-loop approximation (\ref{GammaPertExp}) for the 1PI vertex
$\hat \Gamma$ valid for generic momenta, energies and transfer, we can
obtain now the Fermi Liquid parameters (vertices) using the definitions
(\ref{two-limits1},\ref{two-limits2},\ref{A},\ref{B}).
%
%
%%%%%%%%%%%%%%%%%%%%%%%%%%%%%%%%%%%%%%%%%%%%%%%%%%%%%%%%%%%%%%%%%%%%%%%%%%%%%%
\section{Temperature dependence of the scattering vertex
         and the Landau function}\label{ScatLan}
%%%%%%%%%%%%%%%%%%%%%%%%%%%%%%%%%%%%%%%%%%%%%%%%%%%%%%%%%%%%%%%%%%%%%%%%%%%%%%
%
%
Taking the appropriate zero transfer limit we obtain from
Eqs.~(\ref{GammaPertExp},\ref{A},\ref{B})
the components of the scattering vertex
\begin{equation}
\label{A2}
A(\theta)=u-u^2+2 \chi(q')u^2-C(q_s)u^2~,
\end{equation}
\begin{equation}
\label{B2}
B(\theta)=-u-u^2+C(q_s)u^2
\end{equation}
with $q'=2k_F \sin \theta$ and $q_s=2k_F \cos \theta$. We absorbed
the factor $\nu_0 \equiv m/2\pi$ into the definition of coupling via
\begin{equation}
\label{usm}
 u \equiv \nu_0 U_0~.
\end{equation}
Definitions and detailed calculations of the functions involved in the 
equations above are given in
Appendices \ref{ApA} and \ref{ApB}.
For the components of the Landau function we obtain in the same way
\begin{equation}
\label{F2}
F(\theta)=u+2 \chi(q')u^2-C(q_s)u^2
\end{equation}
\begin{equation}
\label{G22}
G(\theta)=-u+C(q_s)u^2
\end{equation}
Note that while calculating the FL vertices (\ref{A},\ref{B})
at the one-loop level
we can put $\nu_R=\nu_0$ (i.e., $m^{\ast}=m$), $Z=1$, 
and  since $L(\Omega, 0)=0$, the ZS contribution to $F$, $G$ is 
zero in the limit (\ref{two-limits2}).

Using the results (\ref{chi0app},\ref{chi1app},\ref{C0F},\ref{C1F})
of the Appendices we obtain
for the Fourier components of the two vertices:
\begin{equation}
\label{A0F}
A_0=u-u^2 \Big( \ln2\lambda -1+{\pi^2 \over 96}T_0^2 \Big)
\end{equation}
\begin{equation}
\label{F0F}
F_0=u-u^2 \Big( \ln2\lambda -2+{\pi^2 \over 96}T_0^2 \Big)
\end{equation}
\begin{equation}
\label{B0F}
B_0=-u+u^2 \Big( \ln2\lambda -1-{\ln2 \over 2}T_0 \Big)
\end{equation}
\begin{equation}
\label{G0F}
G_0=-u+u^2 \Big( \ln2\lambda -{\ln2 \over 2}T_0 \Big)
\end{equation}
\begin{equation}
\label{AF1F}
A_1=F_1=u^2 \Big( \frac12 -{\pi^2 \over 96}T_0^2 \Big)
\end{equation}
\begin{equation}
\label{BG1F}
B_1=G_1=u^2 \Big(- \frac12 +{\ln2 \over 2}T_0 \Big)
\end{equation}
The terms omitted in the above results are ${\cal O}(u^2 T_0^3)$
at most. Within one-loop
accuracy the standard RPA-type relationships between the FL
parameters $A_n=F_n/(1+F_n)$, $B_n=G_n/(1+G_n)$ hold.

Note that taken separately, each contribution of the
ZS'- or BCS bubble has a leading linear temperature-dependent term
in the first two Fourier components of the vertices [cf.
Eqs.(\ref{chi0app},\ref{chi1app}) and Eqs.(\ref{C0F},\ref{C1F}),
resp.]. However, a cancellation of such terms coming from
two graphs occurs in the ``charge sector'' (i.e., in $A$, $F$ components),
while the linear T-terms survive in the ``spin sector''
($B$, $G$ components).

To  understand where the temperature dependence of the vertices comes 
from, note that  $\chi(q')$ as a function of $q'$ has some
``temperature smearing'' near $q' \sim 2k_F$, otherwise it
is virtually indistinguishable from its zero-temperature limit (\ref{LT0}),
provided the condition (\ref{T0}) applies. So the temperature dependence of 
the ZS' (exchange) contribution to the Fourier components of the vertices 
comes from integrating of the function $\chi$ around the ``effective 
transfer'' $q' \sim 2k_F$ in the ZS' graph, i.e., when incoming momenta
${\bf k}_1 \sim -{\bf k}_2$, while $|{\bf k}_1|=|{\bf k}_2|=k_F$.
Notice that the transfer $q$ in the proper sense
(\ref{GammaTr},\ref{Transfer}) is zero. This source of temperature 
dependence was not considered by Pethick and Carneiro, for example.
In the same vein, one can see from
Eq.(\ref{CqF}) that the temperature dependence of the BCS contribution 
comes from regions of small $q_s$,
i.e., again when ${\bf k}_1 \sim -{\bf k}_2$. The coefficients
in the temperature expansions of the Fourier components of $\chi$ and $C$
are such [cf. Eqs.(\ref{chi0app},\ref{chi1app},\ref{C0F},\ref{C1F})] that the
cancellation occurs only for the linear terms in the charge components of the
vertices, while the higher order terms in the temperature survive. Since to
the best of our knowledge there is no particular reason for this
cancellation of the linear terms to happen, we expect that linear $T$-term
of the charge component(s) would come out of a more realistic calculations of
the FL parameters.

Let us elaborate on the last point a bit more. First notice that in the
perturbative approximation (\ref{GammaPertExp}) for the vertex in the model
with a constant couplings $U_0$ (bare vertex), all the three one-loop terms
have the same factor $U^2_0$ in front of the bubble contributions
$L,~{\Bbb C}$. Had these coupling factors been different
for each of the three graph, the cancellation of the temperature
corrections would not have happened. On the other hand, even the simplest
approximation (\ref{GammaPertExp}) shows that fermionic
interactions drive any bare {\it coupling constant} towards a {\it coupling
function} of, generally speaking, three momenta [cf. Eq.(\ref{GammaTr})].
More advanced calculations of the FL parameters, like, e.g.,  RG, 
would reproduce such development of the momentum dependence in couplings 
in the regime of effective action (\ref{Lamea}). This would prevent the
cancellations between different terms, e.g., between the 
one-loop ZS' and BCS contributions,
which we have obtained in the simple perturbational framework.\cite{note1}

Another artifact of the naive perturbative calculation of the vertex is its
ultraviolet divergence coming from the BCS bubble. The $\ln \lambda$-term
makes our perturbation theory applicable only if the interaction is small
enough that the arbitrary, generally speaking, ultaviolet cutoff $\lambda$ can
be chosen to satisfy $1 \ll \lambda < \frac12 {\rm e}^{1/ u},~(u<1)$. However, 
we should not worry too much about this issue, since more sophisticated 
calculations based, e.g., on summations of the diagrammatic series or an RG would
result in an effective screening of the repulsive interaction in the BCS 
channel. See, e.g., Refs.[\onlinecite{AGD,Lifshitz80,Shankar94}].
(Although providing  the possibility of the {\it infrared} instability (divergence)
at exponentially low temperatures (much lower than, say, we are working at)
through the Kohn-Luttinger mechanism.\cite{KL})

The linear dependence of the leading temperature corrections to
the FL parameters seems not to be sensitive to the actual form of
the one-particle spectrum. The same dependence (apart from 
model-sensitive prefactors) was obtained in the
previous RG analysis of the {\it effective action} model for 2D
spinless fermions with a linearized spectrum.\cite{GD98} In that
study the integration of the ``effective'' transfer in the
exchange (ZS') graph contribution to the RG flow was also
resulting in the linear $T$-term in the vertex corrections.
According to Ref.~[\onlinecite{BK97}] such temperature behavior
can be understood  from dimensional arguments.
%
%
%%%%%%%%%%%%%%%%%%%%%%%%%%%%%%%%%%%%%%%%%%%%%%%%%%%%%%%%%%%%%%%%%%%%%%%%%%%%%%
\section{Ward Identities}\label{WIsec}
%%%%%%%%%%%%%%%%%%%%%%%%%%%%%%%%%%%%%%%%%%%%%%%%%%%%%%%%%%%%%%%%%%%%%%%%%%%%%%
%
%
In order to derive the Ward identities, we followed  closely
the methodology of Ref.[\onlinecite{Lifshitz80}]
adapted to the path-integral technique according to 
standard approaches.\cite{Zinn99} The partition function is given by
the path integral (\ref{Z}). In coordinate-imaginary-time space
the free part of the action is
\begin{equation}
\label{S0ds}
S_0 =  \int d {\it  X}
\bar\psi_{\alpha}({\it X})  \left[
-{\partial \over \partial \tau}+\mu + { {\mathbf{\Delta}} \over 2m}
\right]\psi_{\alpha}({\it X})
\end{equation}
We consider pair interactions which preserve the total spin and  number
of particles
\begin{equation}
\label{PInt}
S_{\rm int}=\int_{({\it X}_1,{\it X}_2)}
\bar\psi_{\alpha}({\it X}_1) \psi_{\alpha}({\it X}_1)
U({\bf x}_1-{\bf x}_2) \delta(\tau_1-\tau_2)
\bar\psi_{\beta}({\it X}_2) \psi_{\beta}({\it X}_2)
\end{equation}
Here
$\int_{({\it X})} \equiv \int_0^{\beta} d \tau \int d {\bf x}$.
Notice that interaction (\ref{SintHub}) is a special case of (\ref{PInt}).
Then the following {\it charge Ward identity}
\begin{equation}
\label{WICs}
{ G^{-1}({\bf 1})-G^{-1}({\bf 1}-{\cal Q}) \over
 i\Omega-{{\bf q}\over 2m}(2{\bf k}_1-{\bf q}) }=
1 - {1 \over N} \int_{({\cal K})}
{ i\Omega-{{\bf q}\over 2m}(2{\bf k}+{\bf q}) \over
 i\Omega-{{\bf q}\over 2m}(2{\bf k}_1-{\bf q}) }
\Gamma^{\alpha \beta}_{\alpha \beta}
({\bf 1},{\cal K};-{\cal Q}) G({\cal K})G({\cal K}+{\cal Q})
\end{equation}
and the {\it spin  Ward identity}
\begin{equation}
\label{WISUNs}
{ G^{-1}({\bf 1})-G^{-1}({\bf 1}-{\cal Q}) \over
 i\Omega-{{\bf q}\over 2m}(2{\bf k}_1-{\bf q}) }=
1-
{1 \over 2(N^2-1)} \int_{({\cal K})}
{ i\Omega-{{\bf q}\over 2m}(2{\bf k}+{\bf q}) \over
 i\Omega-{{\bf q}\over 2m}(2{\bf k}_1-{\bf q}) }
 \lambda^a_{\gamma \varepsilon}
\Gamma^{\alpha \varepsilon}_{\beta \gamma}
({\bf 1},{\cal K};-{\cal Q})
 \lambda^a_{\beta \alpha}
G({\cal K})G({\cal K}+{\cal Q})
\end{equation}
can be derived. 
From  (\ref{Vrep},\ref{LamNor}) one can find
\begin{mathletters}
\label{WIVcs}
\begin{eqnarray}
\Gamma^{\alpha \beta}_{\alpha \beta} &=& -N \Gamma_{\rm ch} \\
\lambda^a_{\gamma \varepsilon}
\Gamma^{\alpha \varepsilon}_{\beta \gamma}
 \lambda^a_{\beta \alpha} &=& -2(N^2-1)\Gamma_{\rm sp} 
\end{eqnarray}
\end{mathletters}
These equations are helpful in order to unclutter the r.h.s of the Ward
identities.
A useful consequence of the Ward identities (\ref{WICs},\ref{WISUNs}) is the
the identity:
\begin{mathletters}
\label{WISCR}
\begin{eqnarray}
\int_{({\cal K})}
 \big[ \Gamma_{\rm ch}({\bf 1},{\cal K};{\cal Q})-
       \Gamma_{\rm sp}({\bf 1},{\cal K};{\cal Q}) \big]
 G({\cal K})G({\cal K}-{\cal Q})
{ i\Omega-{{\bf q}\over 2m}(2{\bf k}-{\bf q}) \over
 i\Omega-{{\bf q}\over 2m}(2{\bf k}_1+{\bf q}) } &=& 0 ~~~
\Longleftrightarrow \\
\int_{({\cal K})}
 \big[ \Gamma^S({\bf 1},{\cal K};{\cal Q})-
       \Gamma^A({\bf 1},{\cal K};{\cal Q}) \big]
 G({\cal K})G({\cal K}-{\cal Q})
{ i\Omega-{{\bf q}\over 2m}(2{\bf k}-{\bf q}) \over
 i\Omega-{{\bf q}\over 2m}(2{\bf k}_1+{\bf q}) } &=& 0
\end{eqnarray}
\end{mathletters}
The above relationships are non-trivial, since
$\Gamma_{\rm ch}({\bf 1},{\cal K};{\cal Q}) \neq \Gamma_{\rm sp}({\bf
1},{\cal K};{\cal Q})$ $[\Gamma^S({\bf 1},{\cal K};{\cal Q}) \neq
\Gamma^A({\bf 1},{\cal K};{\cal Q})]$. There is also another Ward 
identity\cite{Lifshitz80}
\begin{equation}
\label{WIGal}
{\bf k}_1  {\partial G^{-1}({\bf 1}) \over \partial i \Omega}=
{\bf k}_1 +
\int_{({\bf 2})} {\bf k}_2\Gamma_{\rm ch}({\bf 1},{\bf 2};-i \Omega)
 G({\bf 2})G({\bf 2}+ i \Omega) \Big\vert_{i \Omega \to 0}
\end{equation}
which follows from the Galilean invariance.
%
%
%%%%%%%%%%%%%%%%%%%%%%%%%%%%%%%%%%%%%%%%%%%%%%%%%%%%%%%%%%%%%%%%%%%%%%%%%%%%%%
\section{Effective mass }\label{efmass}
%%%%%%%%%%%%%%%%%%%%%%%%%%%%%%%%%%%%%%%%%%%%%%%%%%%%%%%%%%%%%%%%%%%%%%%%%%%%%%
%
%
%%%%%%%%%%%%%%%%%%%%%%%%%%%%%%%%%%%%%%%%%%%%%%%%%%%%%%%%%%%%%%%%%%%%%%%%%%%%%%
\subsection{FLT calculation}\label{FLTemass}
In this subsection we calculate the effective mass using the Ward identities.
The self-energy is defined such that
\begin{equation}
\label{Sigma}
 G^{-1}({\bf 1})= G^{-1}_0({\bf 1})-\Sigma({\bf 1})
\end{equation}
From now on we assume $G_0({\bf 1})$ in the above equation to be not
the exact bare Green's function of the non-interacting system, but
that with the shifted chemical potential. We gauge the shift to be such
that it removes all tadpole insertions to the Green's functions in
diagrammatics.

The effective mass $m^{\ast}$ is given by the following equation
\begin{equation}
\label{mstara}
{m^{\ast} \over m}={ 1-
{\partial \Sigma({\bf 1}) \over \partial i \Omega}
\over
1+{m \over k_F} {\partial \Sigma({\bf 1}) \over \partial {\bf k} }
{{\bf k}_1  \over k_F}  }
\Bigg\vert^{i \omega_1 \leadsto 0}_{{\bf k}_1 \in S_F}
\end{equation}
Since we are working with the finite-temperature Matsubara Green's
functions and $\omega_1$ is a Fermionic Matsubara frequency, we
understand $i \omega_1 \leadsto 0$ in the sense that
the zero-frequency limit will be taken after the appropriate
analytical continuation of the final result for $m^{\ast}({\bf
1})$. In the second order of the perturbation theory only the
``sunrise'' self-energy diagram contributes to the mass
renormalization. In this  approximation
\begin{equation}
\label{mstarappr}
{m^{\ast} \over m}=1-
\Bigg[
{\partial \Sigma({\bf 1}) \over \partial i \Omega}
+
{m \over k_F} {\partial \Sigma({\bf 1}) \over \partial {\bf k} }
{{\bf k}_1  \over k_F}
\Bigg]
\Bigg\vert^{i \omega_1 \leadsto 0}_{{\bf k}_1 \in S_F}
+{\cal O}(U^3)
\end{equation}
The charge Ward identity can be written as
\begin{equation}
\label{WISigk}
{\partial \Sigma({\bf 1}) \over \partial {\bf k} }=
\int_{({\bf 2})} { {\bf k}_2 \over m} \Gamma_{\rm ch}
({\bf 1},{\bf 2};- {\bf q})
 G({\bf 2})G({\bf 2}+ {\bf q}) \Big\vert_{{\bf q} \to 0}
\end{equation}
while the Galilean-invariance Ward identity (\ref{WIGal}) reads
\begin{equation}
\label{WISigGal}
 {\partial \Sigma({\bf 1}) \over \partial i \Omega}= -
\int_{({\bf 2})}{ {\bf k}_1 {\bf k}_2 \over k_1^2}
\Gamma_{\rm ch}({\bf 1},{\bf 2};-i \Omega)
 G({\bf 2})G({\bf 2}+ i \Omega) \Big\vert_{i \Omega \to 0}
\end{equation}
Combining the three above equations together, we get
\begin{equation}
\label{mstarWIap}
{m^{\ast} \over m}=1
+\int_{({\bf 2})}{ {\bf k}_1 {\bf k}_2 \over k_F^2}
\Big[
\Gamma_{\rm ch}({\bf 1},{\bf 2};-i \Omega)
 G({\bf 2})G({\bf 2}+ i \Omega) \Big\vert_{i \Omega \to 0}
-
\Gamma_{\rm ch}({\bf 1},{\bf 2};- {\bf q})
 G({\bf 2})G({\bf 2}+ {\bf q}) \Big\vert_{{\bf q} \to 0}
\Big]
\Big\vert^{i \omega_1 \leadsto 0}_{{\bf k}_1 \in S_F}
\end{equation}
Since the above equation has the accuracy  ${\cal O}(U^2)$, we can use
the one-loop approximation for the vertex, while the Green's functions
can be approximated by $G_0$.

For the the contact interaction (\ref{SintHub}) ($N=2$)
the vertex's one-loop approximation (\ref{GammaPertExp}) gives
\begin{equation}
\label{calA}
\Gamma_{\rm ch}({\bf 1},{\bf 2}; -{\cal Q})
=
U_0+L({\cal Q})U_0^2-2 L({\bf 1}-{\bf 2}-{\cal Q})U_0^2-
{\Bbb C}({\bf 1}+{\bf 2})U_0^2
\end{equation}
It is easy to verify that only the third and the fourth terms in
the above expression (coming from ZS' and BCS graphs, resp.) give
non-zero contributions to $m^{\ast}$. Plugging in the formulas for
the  ZS' and BCS graph contributions and carrying out some
manipulations, we obtain (recall that $i \omega_1 \leadsto
0,~{\bf k}_1 \in S_F $)
\begin{equation}
\label{msWIexpl}
{m^{\ast} \over m}=1+ \frac14 U_0^2
\int_{({\bf k}_2,{\bf k}_3)}{ {\bf k}_1 {\bf k}_2 \over k_F^2}
\Big\{
{ \tanh(\frac12 \beta \xi_{{\bf k}_3}) -
\tanh(\frac12 \beta \xi_{{\bf k}_3+{\bf k}_2-{\bf k}_1}) \over
-i \omega_1+\xi_{{\bf k}_2}+\xi_{{\bf k}_3}
-\xi_{{\bf k}_3+{\bf k}_2-{\bf k}_1}  }
-{ \tanh(\frac12 \beta \xi_{{\bf k}_3}) \over
-i \omega_1-\xi_{{\bf k}_2}+\xi_{{\bf k}_3}+
\xi_{{\bf k}_3-{\bf k}_2-{\bf k}_1}  }
\Big\}
 {\beta \over \cosh^2( \frac12 \beta \xi_{{\bf k}_2} )}
\end{equation}
Note that without applying the Ward identities, the above equation
can be derived  directly from Eq.~(\ref{mstarappr}) where the
self-energy is given by the sunrise diagram.  However, it takes
much more tedious and lengthy calculations to verify the
cancellation of other terms appearing at intermediate steps of the
analysis.

Before proceeding further with the calculations, let us first comment
on result (\ref{msWIexpl}). In the the zero-temperature limit
$\beta / \cosh^2(\frac12 \beta \xi_{{\bf k}_2} ) \propto
\delta(\xi_{{\bf k}_2})$ enforces ${\bf k}_2$ to lie on the Fermi
surface, while the integration over the angle
$\widehat{{\bf k}_1 {\bf k}_2}$ is equivalent to the calculation of the
first Fourier harmonic of the expression in the curly brackets. The latter
we have calculated already, and it is the sum of the ZS' and BCS
contributions to the FL vertex. (Recall that the first order term and
the ZS graph do not contribute to the first Fourier component of the
vertex.) Then Eq.~(\ref{msWIexpl}) in the the zero-temperature limit
readily recovers the standard result of the
FLT:\cite{Landau,AGD,Lifshitz80}
\begin{equation}
\label{msFLT}
{m^{\ast}(T=0) \over m}=1+F_1(T=0)
%\approx 1+\frac12 u^2
\end{equation}
A straightforward extension of Eq.~(\ref{msFLT}) to finite temperatures
like $m^{\ast}(T)/ m=1+F_1(T)$ is
not valid, since according to Eq.~(\ref{msWIexpl}) $m^{\ast}(T)$ contains
an extra contribution from the ``off-shell'' integration over
$k_2~(\xi_{{\bf k}_2})$ normal to the Fermi surface, albeit the factor
$\beta / \cosh^2(\frac12 \beta \xi_{{\bf k}_2} )$ makes this contribution
well-localized near the Fermi surface. In other words the vertex entering
the r.h.s. of Eq.~(\ref{msWIexpl}) is not exactly the FL vertex $F(T)$ as
it is defined in the FLT (cf. definitions in Sec.~\ref{Vertex}),
since one of its  momenta (namely, ${\bf k}_2$)
is not confined to the Fermi surface.

The calculation of the temperature expansion of the effective mass
$m^{\ast}$ is quite involved.\cite{note3}  We did not go beyond the
linear-temperature
terms. Evaluation of the contributions of two graphs to $m^{\ast}$
denoted as
\begin{equation}
\label{msdiag}
{m^{\ast} \over m}=1+u^2( {\cal M}_{ZS'}+{\cal M}_{BCS} )
\end{equation}
gives
\begin{mathletters}
\label{mterms}
\begin{eqnarray}
{\cal M}_{ZS'} &=& \frac14(1+\ln2)T_0 +{\cal O}(T_0^2)\\
{\cal M}_{BCS} &=& \frac12 -\frac14(1+\ln2)T_0 +{\cal O}(e^{-1/T_0})
\end{eqnarray}
\end{mathletters}
Thus we get
\begin{equation}
\label{msFin}
{m^{\ast} \over m}=1+ \frac12 u^2 +{\cal O}(u^2T_0^2)
\end{equation}
(Recall that dimensionless coupling and temperature are
introduced by (\ref{usm}) and (\ref{T0}), resp.) In a close
analogy with the cancellation of the linear-temperature terms in
the Fourier components of the FL vertices $A$ and $F$, here the
cancellation occurs between additive linear-$T$ corrections coming
from both ``on-shell'' (i.e., linear $T$-term coming from the 
$2k_F$-contribution to the vertex) and ``off-shell'' (i.e., small-momentum
contribution) integrations in two
diagrams. Moreover, the ``on-shell'' (``off-shell'') $T$-term of
the ZS' graph cancels  the ``on-shell'' (``off-shell'') $T$-term
of the BCS graph, correspondingly.
%
%
%
%%%%%%%%%%%%%%%%%%%%%%%%%%%%%%%%%%%%%%%%%%%%%%%%%%%%%%%%%%%%%%%%%%%%%%%%%%%%%%
\subsection{Alternative Calculation}\label{altm}
In this subection we present an alternative direct evaluation of the leading
temperature correction.  The advantage of this evaluation is that it 
applies also to non-Galilean invariant situations and helps
to more clearly establish the relation to previous 
studies. It also indicates possible generalizations.

We begin from the explicit expression for the self-energy correction given by 
the ``sunrise'' diagram
\begin{equation}
\Sigma ({\cal P})=U_0^{2}\int_{\cal Q, K}
G({\cal P+Q+K})G(-{\cal K})G({\cal Q})
\end{equation}
This choice of variables is convenient
because the $T-$linear contributions to the effective mass  will be seen to
arise either from regions where ${\bf q}+{\bf k}$ is
small or from regions where it has magnitude $2p_{F}$. The bare electron
Green function $G$ is given by (\ref{G0def}).
Evaluation of the Matusbara sums leads to 
\begin{equation}
\Sigma (p,i\omega )=U_0^{2}\int_{\bf k,q} 
 \frac{[n_{\circ}(\xi _{k})-n_{\circ}(\xi _{q})][b(\xi
_{q}-\xi _{k})+n_{\circ}(\xi_{\bf p+q+k})]}{i\omega +\xi
_{q}-\xi _{k}-\xi_{\bf p+k+q}}
\end{equation}
and by using $[n_{\circ}(\xi
_{k})-n_{\circ}(\xi _{q})][b(\xi _{q}-\xi
_{k})]=n_{\circ}(\xi _{q})[1-n_{\circ}(\xi _{k})]$ 
($n_{\circ}$ and $b$ are the Fermi-Dirac and Bose-Einstein
distribution functions, resp.), and relabelling variables we obtain 
\begin{equation}
\Sigma (p,i\omega )=U_0^{2}\int_{\bf k,q}
\left( \frac{n_{\circ}(\xi _{q})}{i\omega +\xi
_{q}-\xi _{k}-\xi _{\bf p+k+q}}-n_{\circ}(\xi _{q})n_{\circ}(\xi
_{k})J\right) 
\end{equation}
with $J=2/(i\omega +\xi _{q}-\xi _{k}-\xi
_{\bf p+k+q})+1/(-i\omega +\xi _{q}+\xi _{k}-\xi
_{\bf p+k+q}).$

Now the first term in the expression for $\Sigma $ cannot give rise to a
term of order $T$ because the integral of $\xi _{k}$ is not confined
to the region near the Fermi surface. Indeed, differentiating this term on
temperature leads to a numerator containing the factor $\xi _{q}$.
Because the $\xi _{k}$ integral is not confined to the region near
the Fermi surface, obtaining a nonzero value for the $\xi _{q}$
integral would require an additional factor of $\xi _{q}$ divided by
a scale of the order of $E_F$, meaning that the integral would be
of order $T^{2}$ as in the usual Sommerfeld expansion. 

We therefore focus on possible T-linear contributions from the second term.  
To obtain these we will make the assumption, to be justified a posteriori, 
that all energies $(\xi_{q}, \xi_{k}, \xi_{p})$ are near zero. We 
parameterize the ${\bf q}$, ${\bf k}$ integrals by $\xi_{k,q}$ 
and angles.  We choose $\theta _{1}$ as the angle between ${\bf p}$ and 
${\bf q}+{\bf k}$ and $\theta
_{2}$ as the angle between ${\bf q}$ and ${\bf k}$.
For a spherical Fermi surface we have 
\begin{equation}
\xi _{\bf p+k+q}=A+B\cos (\theta _{1})
\end{equation}
with, (up to first order in deviations from the Fermi surface) 
\begin{eqnarray}
A &=&[4E_{F}+\xi _{q}+\xi _{k}]\cos ^{2}(\theta
_{2}/2)+\xi _{p} \\
B &=&[4E_{F}+2\xi _{p}+\xi _{q}+\xi _{k}]\cos
(\theta _{2}/2)
\end{eqnarray}

We may now do the integral over $\theta _{1}$ obtaining 
\begin{equation}
\Sigma (p,i\omega )=-{u^2 \over 2\pi} \int d\xi
_{q}d\xi _{k}d\theta _{2}n_{\circ}(\xi _{k})n_{\circ}(\xi _{q})K
\end{equation}
with 
\begin{equation}
K=\frac{2sgn(C_{1})}{%
%TCIMACRO{\func{Re}}%
%BeginExpansion
\mathop{\rm Re}%
%EndExpansion
[\sqrt{C_{1}^{2}-B^{2}}]}+\frac{sgn(C_{2})}{%
%TCIMACRO{\func{Re}}%
%BeginExpansion
\mathop{\rm Re}%
%EndExpansion
[\sqrt{C_{2}^{2}-B^{2}}]}
\end{equation}
and 
\begin{eqnarray}
C_{1} &=&\omega +\xi _{q}-\xi _{k}-\xi
_{p}-[4E_{F}+\xi _{q}+\xi _{k}]\cos ^{2}(\theta _{2}/2) \\
C_{2} &=&-\omega +\xi _{q}+\xi _{k}-\xi
_{p}-[4E_{F}+\xi _{q}+\xi _{k}]\cos ^{2}(\theta _{2}/2)
\end{eqnarray}
The crucial point is now that (after analytic continuation in $\omega $) the
region of phase space in which the square root is real is very small, and in
particular corresponds to $\theta _{2}\approx  0$ or $\theta _{2} \approx
\pm \pi $, as well as to small values of the energies, thus justifying the 
assumption made above. Let us consider
these two regions separately.
%%%%%%%%%%%%%%%%%%%%%%%%%%%%%%%%%%%%%%%%%%%%%%%%%%%%%%%%%%%%%%%%%%%%%%%%%%%%

$\bullet$ $\theta _{2} \approx \pi $:
These angles correspond to processes in which a small total momentum is
transferred to the electron, i.e. to the processes which are usually
considered\cite{Bedell93,CP77} to give rise to anomalous  terms in $C/T$, 
etc. In this regime we write $\theta _{2}=\pm \pi -y$; to acount for 
$\pm \pi $, y may have either sign. To leading order we have 
\begin{eqnarray}
C_{1} &=&\omega +\xi _{q}-\xi _{k}-\xi _{p} \\
C_{2} &=&-\omega +\xi _{q}+\xi _{k}-\xi _{p} \\
B &=&2E_{F}y
\end{eqnarray}
Note that if all the $\xi $ are small, then $\theta _{2}$ is confined
to small angles, where the approximation is accurate. Performing the angle
integral gives 
\begin{equation}
\Sigma ^{small}(p,\omega )=-{u^2 \over 4E_F}\int
d\xi _{q}d\xi _{k}n_{\circ}(\xi _{k})n_{\circ}(\xi
_{q})L^{small}
\end{equation}
with 
\begin{equation}
L^{small}=2sgn(\omega +\xi _{q}-\xi _{k}-\xi
_{p})+sgn(-\omega +\xi _{q}+\xi _{k}-\xi _{p})
\end{equation}
The remainder of the evaluation is simple: to obtain the quasiparticle
velocity renormalization we must take the sum of the $\omega $ and $%
\xi _{p}$ derivatives; thus the first term does not contribute at
all while the second term gives $-2\delta (\xi _{q}+\xi
_{k}) $. for the $\omega $ derivative and the same for the $\xi _{p}$
derivative. Substituting this into the integrals and evaluating gives 
\begin{equation}
\frac{m^{\ast }}{m}|_{small}={u^{2} \over E_{F}} \int
d\xi _{k}n_{\circ}(\xi _{k})n_{\circ}(-\xi _{k})={u^2T \over E_F}
\end{equation}

%%%%%%%%%%%%%%%%%%%%%%%%%%%%%%%%%%%%%%%%%%%%%%%%%%%%%%%%%%%%%%%%%%%%%%%%%%%%%%%%
$\bullet$ $\theta _{2} \approx 0$:
These angles correspond to processes in which a momentum near $2p_{F}$ is
tranferred to the electrons. \ In this regime we have $%
sgn(C_{1})=sgn(C_{2})=-1.$ Further, as both $C_{1,2}$ and $B$ are of order $%
E_{F}$ we write $C^{2}-B^{2}=(C+B)\ast (C-B)=-8E_{F}\ast (C+B)$. \ We have 
\begin{eqnarray}
C_{1}+B &=&\omega -\xi _{k}+\xi _{q}+\xi _{p}+\frac{1%
}{2}E_{F}\theta _{2}^{2} \\
C_{2}+B &=&-\omega +\xi _{p}+\xi _{q}+\xi _{k}+\frac{%
1}{2}E_{F}\theta _{2}^{2}
\end{eqnarray}
Again the $\theta _{2}$ integral may be done, yielding 
\begin{equation}
\Sigma ^{2p_{F}}(p,\omega )=- {u^2 \over 4 E_F} \int
d\xi _{q}d\xi _{k}n_{\circ}(\xi _{k})n_{\circ}(\xi
_{q})L^{2p_{F}}
\end{equation}
with 
\begin{equation}
L^{2p_{F}}=2\Theta (\omega -\xi _{k}+\xi _{q}+\xi
_{p})+\Theta (-\omega +\xi _{p}+\xi _{q}+\xi _{k})
\end{equation}
As above, one of the two terms give no contribution; the other has a factor
of two but when differentiated gives only $\delta $ not $2\delta $. $\ $\
The sign is opposite. Thus this term cancels the previous term, leading to
no T-linear term in the effective  mass enhancement. 

An almost identical 
calculation of the order $u^{2}$ term in the free energy shows that there 
is no T-linear term in the specific heat coefficient. 

As we see from these calculations, the coefficients of the linear 
$T$-contributions to the effective mass coming from small and nearly
$2p_F$ processes are model-dependent. Any model modifications, e.g., 
momentum dependent interaction or non-circular Fermi surface, changes
these coefficients, eliminating the cancellation and 
resulting in  apperance of the linear $T$-term
in the effective mass. The same conclusion applies to the specific heat
coefficient. This agrees with the recent experimental data on liquid 
$~^3{\rm He}$ films.\cite{Misawa99}
%
%
%%%%%%%%%%%%%%%%%%%%%%%%%%%%%%%%%%%%%%%%%%%%%%%%%%%%%%%%%%%%%%%%%%%%%%%%%%%%%%
\section{Density-density correlation function }\label{compr}
%%%%%%%%%%%%%%%%%%%%%%%%%%%%%%%%%%%%%%%%%%%%%%%%%%%%%%%%%%%%%%%%%%%%%%%%%%%%%%
%
%
In this section we
consider the density-density correlation function $\varkappa({\cal Q})$,
defined as follows:
\begin{equation}
\label{DR}
\delta^{(d+1)}(0)\varkappa({\cal Q}) \equiv \langle \tilde{\rho}({\cal Q})
\tilde{\rho}^{\dag}({\cal Q})\rangle
\end{equation}
where the density operator $\rho({\cal Q})$ is
\begin{equation}
\label{DOP}
\rho({\cal Q})=\int_{({\bf 1})}\bar\psi_{\alpha}({\bf 1})
\psi_{\alpha}({\bf1}+{\cal Q})
\end{equation}
and $\tilde{\rho}({\cal Q})=\rho({\cal Q})-\langle \rho({\cal Q}) \rangle $
stands for the density fluctuation.
Using the definition for the 1PI four-point vertex (\ref{2PV}), one obtains
\begin{equation}
\label{kapC}
\varkappa({\cal Q})
=-N \Big[ \int_{({\bf 1})} G({\bf 1})G({\bf 1}-{\cal Q})
+ \int_{({\bf 1},{\bf 2})} G({\bf 1})G({\bf 1}-{\cal Q})
\Gamma_{\rm ch}({\bf 1},{\bf 2};-{\cal Q})
 G({\bf 2})G({\bf 2}+{\cal Q}) \Big]
\end{equation}
The correlation function has the properties
\begin{mathletters}
\label{kapProp}
\begin{eqnarray}
\varkappa({\bf q}=0,\Omega \neq 0)=0 \\
\varkappa({\cal Q})= \varkappa(-{\cal Q})
\end{eqnarray}
\end{mathletters}
To derive (\ref{kapProp}a) it is enough to re-write
the charge Ward identity as
\begin{eqnarray}
\label{WI}
G({\bf 1}-{\cal Q}) G({\bf 1}) & = &
{G({\bf 1}-{\cal Q}) - G({\bf 1}) \over
i \Omega - {{\bf q}(2{\bf k}_1-{\bf q}) \over 2m} } \nonumber \\
&~ & - \int_{({\bf 2})}
{i \Omega - {{\bf q}(2{\bf k}_2+{\bf q}) \over 2m}  \over
i \Omega - {{\bf q}(2{\bf k}_1-{\bf q}) \over 2m} }
G({\bf 1}) G({\bf 1}-{\cal Q})
\Gamma_{\rm ch}({\bf 1},{\bf 2};-{\cal Q})
 G({\bf 2}) G({\bf 2}+{\cal Q})
\end{eqnarray}
Using the above expression for the first term in the brackets on the
r.h.s. of Eq.~(\ref{kapC}), one can easily establish Eq.~(\ref{kapProp}a).
As to the second Eq.~(\ref{kapProp}b), it can be proved with the use of
Eqs.(\ref{versym}).

In order to find the compressibility of the system we need to
calculate the density function in the non-trivial (static) zero
transfer limit, which  we denote $\varkappa^q \equiv
\varkappa({\bf q} \to 0,\Omega =0)$. To do it in the second order
over interaction by a naive perturbation theory, one needs to
take into account the sunrise diagram for the self-energy in the
Dyson equation (\ref{Sigma}) and to use such approximation for
the total Green's function entering the first term in the
brackets on the r.h.s. of Eq.~(\ref{kapC}). For the second term
in that equation one can use the ``bare'' Green's functions [cf.
comment after Eq.~(\ref{Sigma})] and the one-loop approximation
for the vertex. In diagrammatic language this amounts to the 
calculation of a set of three-loop diagrams, which is very 
difficult in practice. The first problem is a
proliferation of terms after doing each intermidiate step of
integration (summation) for a given three-loop diagram. The
second is that some of these terms exhibit IR 
divergencies.\cite{NMetz98} It takes additional efforts to assure 
the final cancellation of those divergencies between different 
terms. As in the effective mass calculation, the use of the Ward 
identities drastically simplifies the problem.
[Cf. comment after Eq.~(\ref{msWIexpl})].

In applying the Ward identities for our calculations we will use
the approaches developed in the classic derivations of
FLT.\cite{Lifshitz80} We denote
\begin{mathletters}
\label{Gsq}
\begin{eqnarray}
 G^2_q({\bf 1}) &\equiv &
 G({\bf 1})G({\bf 1}+{\bf q}) \Big\vert_{{\bf q} \to 0} \\
 G^2_{\Omega}({\bf 1}) &\equiv&
 G({\bf 1})G({\bf 1}+ i \Omega) \Big\vert_{i \Omega \to 0}
\end{eqnarray}
\end{mathletters}
Let us also introduce two quantities $\Delta_R$ and $K$ as
\begin{equation}
\label{dR}
 G^2_q({\bf 1})= G^2_{\Omega}({\bf 1})-\Delta_R({\bf 1})
\end{equation}
\begin{equation}
\label{dAqo}
 \Gamma_{\rm ch}^q({\bf 1},{\bf 2})=
\Gamma_{\rm ch}^{\Omega}({\bf 1},{\bf 2})
-{K}({\bf 1},{\bf 2})
\end{equation}
Then from Eqs.(\ref{kapC},\ref{dAqo}) we have
\begin{equation}
\label{kapq1}
-{ \varkappa^q \over N}= \int_{({\bf 1})} G^2_q({\bf 1})
+ \int_{({\bf 1},{\bf 2})}
G^2_q({\bf 1}) \Gamma_{\rm ch}^q({\bf 1},{\bf 2}) G^2_q({\bf 2})
-\varkappa_2
\end{equation}
where
\begin{equation}
\label{kap2}
\varkappa_2 = \int_{({\bf 1},{\bf 2})}
G^2_q({\bf 1}) {K}({\bf 1},{\bf 2}) G^2_q({\bf 2})
\end{equation}
Eqs.~(\ref{kapC},\ref{kapProp}a) give
\begin{equation}
\label{kapOM}
\int_{({\bf 1})} G^2_{\Omega}({\bf 1})
+ \int_{({\bf 1},{\bf 2})}
G^2_{\Omega}({\bf 1}) \Gamma_{\rm ch}^{\Omega}({\bf 1},{\bf 2})
G^2_{\Omega}({\bf 2})=0
\end{equation}
Subtracting this from Eq.~(\ref{kapq1}) and using (\ref{dR}), we
obtain
\begin{equation}
\label{kapq2}
-{ \varkappa^q \over N}=- \int_{({\bf 1})} \Delta_R({\bf 1})
\Big[ 1+2 \int_{({\bf 2})}
\Gamma_{\rm ch}^{\Omega}({\bf 1},{\bf 2})G^2_{\Omega}({\bf 2}) \Big]
+  \int_{({\bf 1},{\bf 2})} \Delta_R({\bf 1})
\Gamma_{\rm ch}^{\Omega}({\bf 1},{\bf 2}) \Delta_R({\bf 2})
-\varkappa_2
\end{equation}
Note that from the charge Ward identity
\begin{equation}
\label{WIsig}
 {\partial \Sigma({\bf 1}) \over \partial i \Omega}= -
 \int_{({\bf 2})}
\Gamma_{\rm ch}^{\Omega}({\bf 1},{\bf 2})G^2_{\Omega}({\bf 2})
\end{equation}
So far we did not use any approximations, and Eq.~(\ref{kapq2}) is exact.
Now let us simplify it at the level ${\cal O}(U_0^2)$ for the model
with interaction (\ref{SintHub}) $(N=2)$.
From the one-loop approximation for the vertex (\ref{calA})
we see that
\begin{equation}
\label{K1L}
{K}({\bf 1},{\bf 2})={m \over 2 \pi}U^2_0
\end{equation}
Also the Green's functions in Eq.~(\ref{kap2}) can be approximated by
their bare forms, resulting in
\begin{equation}
\label{kap2ap}
\varkappa_2 = {m \over 2 \pi}u^2+{\cal O}(u^3)
\end{equation}
After the standard analytical continuation to real frequencies,
one can find for the parameter
$\Delta_R({\bf 1})$,  ${\bf 1} \equiv ({\bf k}_1,\omega_1)$
\begin{equation}
\label{DeltaRap}
 \Delta_R({\bf 1}) \approx
\Big(1+ {\partial \Sigma({\bf 1}) \over \partial  \omega}-
{m \over k_F} {\partial \Sigma({\bf 1}) \over \partial {\bf k} }
{{\bf k}_1  \over k_F} \Big)
 \Delta({\bf 1})
\end{equation}
where
\begin{equation}
\label{Delta1}
\Delta({\bf 1})=\frac12 {\goth C}_1 \delta(\omega_1-\xi_{{\bf k}_1})~,~~
{\goth C}_1 \equiv {\beta \over 2}
{1 \over \cosh^2( \frac12 \beta \xi_{{\bf k}_1}) }
\end{equation}
Eq.~(\ref{DeltaRap}) is the extension to the finite-temperature
of the well-known result
$\Delta_R({\bf 1},T=0)=
Z^2m^{\ast}/k_F\delta(\omega_1)\delta(k_1-k_F)$ (cf., e.g.,
Ref.~[\onlinecite{Lifshitz80}]), combined with expansion 
of $Z^2m^{\ast}$ to order $u^2$.
Up to the terms ${\cal O}(u^3)$ we have from
Eqs.~(\ref{kapq2},\ref{WIsig},\ref{kap2ap},\ref{DeltaRap})
\begin{equation}
\label{kapq3}
-{ \varkappa^q \over 2}=-{m \over 2 \pi}-{m \over 2 \pi}u^2
+ \int_{({\bf 1})} \Delta({\bf 1})
\Big( {\partial \Sigma({\bf 1}) \over \partial \omega}+
{m \over k_F} {\partial \Sigma({\bf 1}) \over \partial {\bf k} }
{{\bf k}_1  \over k_F} \Big)
+  \int_{({\bf 1},{\bf 2})} \Delta({\bf 1})
\Gamma_{\rm ch}^{\Omega}({\bf 1},{\bf 2}) \Delta({\bf 2})
\end{equation}
As we have shown before for the calculation of the effective
mass, the first integral on the r.h.s. of Eq.~(\ref{kapq3}) can
be written via the Ward identities as the second term on the
r.h.s. of Eq.~(\ref{mstarWIap}) (With the only difference that
in Eq.~(\ref{kapq3}) $({\bf k}_1,\omega_1)$ is arbitrary.) Since
two limits of the vertex differ only by a constant term
(\ref{K1L}) which disappears after the integration over the angle
$\widehat{{\bf k}_1 {\bf k}_2}$, we can finally write
\begin{eqnarray}
\label{kapq4}
\varkappa^q   &=& {m \over  \pi} \Big\{   1+u^2 +f_1-f_0 \Big\} \\
f_1 &\equiv&
{2 \pi \over m}  \int_{({\bf 1},{\bf 2})} {{\bf k}_1 {\bf k}_2 \over k^2_1}
\Delta({\bf 1}) \Gamma_{\rm ch}^{\Omega}({\bf 1},{\bf 2})
\Delta({\bf 2}) \nonumber \\
f_0 &\equiv&
{2 \pi \over m}  \int_{({\bf 1},{\bf 2})}
\Delta({\bf 1}) \Gamma_{\rm ch}^{\Omega}({\bf 1},{\bf 2})
\Delta({\bf 2}) \nonumber
\end{eqnarray}
Eqs.~(\ref{kapq4},\ref{calA}) for the response function are manifestly
free of spurious IR divergences and readily recover the standard FLT
results at zero temperature.
We recall  [cf. Eq.~(\ref{A})] that the components
of the scattering vertex and the Landau function are just the appropriate
limits of the vertex $\Gamma_{\rm ch}$ times the  factor $\nu_R Z^2$
$(\nu_R=m^{\ast}/2 \pi)$. The latter
reduces simply to $m / 2\pi$ in our approximation. Then at $T=0$
[cf. Sec.~\ref{efmass} for the explanations of $T \to 0$ limit, and note also
that from Eq.~(\ref{F0F}) $F_0^2=u^2+{\cal O}(u^3)$]
we read off from Eq.~(\ref{kapq4})
\begin{equation}
\label{kapqT0}
\varkappa^q(T=0)= {m \over  \pi}(1+F_0^2+F_1-F_0)
\end{equation}
which is nothing but the expansion to ${\cal O}(u^3)$ of 
2D FLT result (cf., e.g., Ref.~[\onlinecite{Lifshitz80}])
\begin{equation}
\label{kapFLT}
\varkappa^{FLT}(T=0)= {m \over  \pi}{ 1+F_1 \over 1+F_0}
\end{equation}
Using Eq.~(\ref{calA}) for the vertex entering the first integral term
in Eq.~(\ref{kapq4}), we obtain:
\begin{eqnarray}
\label{f1}
f_1 &=& {2 \pi \over m} U_0^2
\int_{({\bf k}_1,{\bf k}_2,{\bf k}_3)}{ {\bf k}_1 {\bf k}_2 \over k_1^2}
{ {\goth C}_1 {\goth C}_2   \over 4}
\Big\{
{ \tanh(\frac12 \beta \xi_{{\bf k}_3}) -
\tanh(\frac12 \beta \xi_{{\bf k}_3+{\bf k}_2-{\bf k}_1}) \over
-\xi_{{\bf k}_1}+\xi_{{\bf k}_2}+\xi_{{\bf k}_3}
-\xi_{{\bf k}_3+{\bf k}_2-{\bf k}_1}  }
-{ \tanh(\frac12 \beta \xi_{{\bf k}_3}) \over
-\xi_{{\bf k}_1} -\xi_{{\bf k}_2}+\xi_{{\bf k}_3}+
\xi_{{\bf k}_3-{\bf k}_2-{\bf k}_1}  }
\Big\} \\
 &\equiv& f_1^{ZS'} + f_1^{BCS} \nonumber
\end{eqnarray}
where we indicated explicitly two contributions to the response function coming
from the ZS' and BCS one-loop corrections of the vertex. For the second integral
on the r.h.s. of Eq.~(\ref{kapq4}) we have
\begin{equation}
\label{f0}
f_0= {m \over 2 \pi} U_0 + f_0^{ZS'} + f_0^{BCS}
\end{equation}
where the first term is the bare vertex contribution to $\varkappa^q$, while
the two other one-loop vertex corrections are given by the same formulas as
in Eq.~(\ref{f1}) with an obvious replacement
${\bf k}_1 {\bf k}_2 / k_1^2 \mapsto 1$.

We were able to analytically calculate the above terms in the leading order
of their temperature dependence.\cite{note3} The results are
\begin{mathletters}
\label{kapcomp}
\begin{eqnarray}
 f_0^{ZS'}&=&u^2(2-\frac12 T_0) \\
 f_0^{BCS}&=&u^2(-\ln2\lambda +\frac12 T_0) \\
 f_1^{ZS'}&=&\frac12 u^2 T_0 \\
 f_1^{BCS}&=&\frac12 u^2(1-T_0)
\end{eqnarray}
\end{mathletters}
up to the order ${\cal O}(u^2T_0^2)$. Note that the BCS
contribution to the term $f_0$ is ultraviolet divergent and we
regularized it in the same way as we did it for that contribution
to the vertex, i.e., by introducing an UV cutoff $\lambda \gg 1$.
[Cf. Sec. \ref{ScatLan}.] Applying Eq.~(\ref{WISigGal}) for the
term $f_1$ while passing  from Eq.~(\ref{kapq3}) to
Eq.~(\ref{kapq4}) may raise some questions from a cautious reader,
since the ultraviolet regularization we used to calculate the BCS
contribution would break the Galilean invariance, which is
indispensable for the derivation of (\ref{WISigGal}). However
$f_1$ is UV-convergent, thus can be calculated in the limit of an
infinite cutoff when the Galilean invariance is preserved.

Combining the above results together, we obtain
\begin{equation}
\label{kapfin}
\varkappa^{q}=
{m \over  \pi}(1-u-\frac12 u^2+u^2\ln2\lambda)+{\cal O}(u^2T_0^2)
\end{equation}
Once again, the leading linear $T$-corrections  which can be traced back to
the ZS'- and BCS-loop contributions to the vertex cancel in each
of the terms $f_0$ and $f_1$ separately.
It is useful to keep in mind that albeit we expressed  $f_1$ and $f_0$ in
terms of the vertex only, these contributions entangle both ``purely vertex''
corrections and self-energy corrections to the response function. The latter
are just expressed in terms of the vertex via the Ward identities.

Let us finally remind\cite{Lifshitz80} that the compressibility
$K=\varkappa^{q}/ n^2$, where $n$ is electron density.
%
%
%%%%%%%%%%%%%%%%%%%%%%%%%%%%%%%%%%%%%%%%%%%%%%%%%%%%%%%%%%%%%%%%%%%%%%%%%%%%%%
\section{Spin-spin correlation function }\label{susc}
%%%%%%%%%%%%%%%%%%%%%%%%%%%%%%%%%%%%%%%%%%%%%%%%%%%%%%%%%%%%%%%%%%%%%%%%%%%%%%
%
%
To find the spin susceptibility in the $SU(N)$ formalism, we consider the
spin (flavor, if $N \neq 2$) correlation function $\chi^{ab}({\bf Q})$,
defined as follows:
\begin{equation}
\label{FR}
\delta^{(d+1)}(0)\chi^{ab}({\cal Q}) \equiv \langle
S^a({\cal Q}) S^{b \dag}({\cal Q})\rangle
\end{equation}
where the flavor density operators
\begin{mathletters}
\begin{eqnarray}
\label{FOP}
S^a({\cal Q}) &=& {{\goth g} \over 2} \int_{({\bf 1})} \lambda^a_{\alpha \beta}
\bar\psi_{\alpha}({\bf 1}) \psi_{\beta}({\bf1}+{\cal Q}) \\
S^{a \dag}({\cal Q}) &=& {{\goth g} \over 2} \int_{({\bf 1})} \lambda^a_{\alpha \beta}
\bar\psi_{\alpha}({\bf 1}) \psi_{\beta}({\bf1}-{\cal Q})
\end{eqnarray}
\end{mathletters}
and ${\goth g}$ is the gyromagnetic ratio. Using Eqs.~(\ref{2PV},\ref{WIVcs})
one can find
\begin{equation}
\label{chiS}
 \chi^{ab}({\cal Q})=
-{{\goth g}^2 \over 2} \delta^{ab} \Big[ \int_{({\bf 1})} G({\bf 1})G({\bf 1}-{\cal Q})
+ \int_{({\bf 1},{\bf 2})} G({\bf 1})G({\bf 1}-{\cal Q})
\Gamma_{\rm sp}({\bf 1},{\bf 2};-{\cal Q})
 G({\bf 2})G({\bf 2}+{\cal Q}) \Big]
\end{equation}
Note that in a paramagnetic state the response is the same along all the
$(N^2-1)$ directions $a$.
Then we follow the same steps as in the previous section, with the only
difference that instead of the charge Ward identity we use the spin identity
(\ref{WISUNs}). The latter can be casted in the form of  Eq.~(\ref{WI})
where $\Gamma_{\rm ch} \mapsto \Gamma_{\rm sp}$. Function
$\chi^{aa}({\cal Q})$ has
the same properties (\ref{kapProp}) as those of the density function.
The uniform susceptibility is given by the static limit
$\chi^q \equiv \chi^{aa}({\bf q}\to 0,\Omega=0)$ (no sum over $a$).
The equation we obtain for
$2\chi^q / {\goth g}^2$ is the same as the r.h.s of Eq.~(\ref{kapq2}), but
$\Gamma_{\rm ch}^{\Omega} \mapsto \Gamma_{\rm sp}^{\Omega}$. 
Then we use the consequence
of the both spin and charge Ward identities [cf.  Eq.~(\ref{WISCR})]:
\begin{equation}
\label{WIsc2}
 \int_{({\bf 2})}
\Gamma_{\rm sp}^{\Omega}({\bf 1},{\bf 2})G^2_{\Omega}({\bf 2})=
 \int_{({\bf 2})}
\Gamma_{\rm ch}^{\Omega}({\bf 1},{\bf 2})G^2_{\Omega}({\bf 2})
\end{equation}
and afterwards proceed as in the previous section.
From the one-loop approximation (\ref{GammaPertExp})
\begin{equation}
\label{calB}
\Gamma_{\rm sp}({\bf 1},{\bf 2};-{\cal Q})=
-U_0+L({\cal Q})U_0^2 +{\Bbb C}({\bf 1}+{\bf 2})U_0^2
\end{equation}
Thus up to the terms ${\cal O}(u^3)$ we obtain
\begin{eqnarray}
\label{chiq4}
\chi^q &=& {m {\goth g}^2 \over 4 \pi} \Big\{ 1+u^2 +f_1- g_0 \Big\} \\
g_0 &\equiv&
{2 \pi \over m}  \int_{({\bf 1},{\bf 2})}
\Delta({\bf 1}) \Gamma_{\rm sp}^{\Omega}({\bf 1},{\bf 2}) \Delta({\bf 2})
\end{eqnarray}
Recalling once again what we said after Eq.~(\ref{kapq4})
and noting that from  Eq.~(\ref{G0F}) $G_0^2=u^2+{\cal O}(u^3)$, we
easily see that in the zero-temperature limit the above equation
gives
\begin{equation}
\label{chiqT0}
\chi^q(T=0)={m {\goth g}^2 \over 4 \pi}(1+G_0^2+F_1-G_0)
\end{equation}
reproducing thus the second-order expansion of 
the 2D FLT result\cite{Lifshitz80}
\begin{equation}
\label{chiFLT}
\chi^{FLT}(T=0)={m {\goth g}^2 \over 4 \pi}{ 1+F_1 \over 1+G_0}
\end{equation}
Luckily we do not have to do new calculations since
\begin{equation}
\label{g0r}
g_0=-u-f_0^{BCS}
\end{equation}
So we find the spin
susceptibility to have a linear leading temperature correction
\begin{equation}
\label{chifin}
\chi^{q} = {m {\goth g}^2 \over 4 \pi}
(1+u+\frac32 u^2-u^2\ln2\lambda +{T_0 \over 2} u^2)+{\cal O}(u^2T_0^2)
\end{equation}
%
%
%%%%%%%%%%%%%%%%%%%%%%%%%%%%%%%%%%%%%%%%%%%%%%%%%%%%%%%%%%%%%%%%%%%%%%%%%%%%%%
\section{Summary and Discussion }\label{Concl}
%%%%%%%%%%%%%%%%%%%%%%%%%%%%%%%%%%%%%%%%%%%%%%%%%%%%%%%%%%%%%%%%%%%%%%%%%%%%%%
%
%
In this paper we have systematically examined the leading temperature 
corrections to FLT in two spatial dimensions. 
We find for the model of a 2D electron gas
with a contact interaction that to order $u^2$ the leading $T$-dependence 
of the FL parameters in the spin sector is linear in temperature, while for 
the parameters in the charge sector and for the effective mass a cancellation 
of the leading  $T$-corrections occurs, and their expansions start from the 
terms quadratic in temperature.  

The particularly interesting result we found is the leading linear 
temperature dependence of the spin susceptibility
\begin{equation}
\label{sus}
{ \chi^q(T) \over  \chi^q(0) } \approx 1+u^2  {T \over E_F}
\end{equation}
According to the perturbative calculations of Belitz {\it et al},\cite{BK97}
the 2D FL susceptibility has a leading linear correction in 
$|{\bf q}|$ at $T=0$
with a positive coefficient which is of the second 
order in interaction, i.e., their result has a structure of 
Eq.~(\ref{sus}). This also agrees with the phenomenology of 
Misawa\cite{Misawa99} and the numerical results.\cite{HT98}

Our results reveal the crucial importance of $2k_F$ processes contributions
into the low-energy parameters. We remind that recovering of
the angular dependence of the FL vertices in the whole region
where the angle varies, involves taking into account transfers varying 
from zero to $2k_F$. We have seen it from the direct bubbles evaluation, 
but the argument is non-perturbative. To appreciate this let us write the
antisymmetry condition for the vertex as
\begin{mathletters}
\label{Cross}
\begin{eqnarray}
\Gamma^{\{ \!\! \mbox{$ \begin{array}{c}
 \mbox{\tiny S} \\[-0.2cm]  \mbox{\tiny A} \end{array} $} \!\! \}}
({\bf k}_1,{\bf k}_2;{\bf q})= \pm
\Gamma^{\{\!\!  \mbox{$ \begin{array}{c}
 \mbox{\tiny S} \\[-0.2cm] \mbox{\tiny A} \end{array} $}\!\! \}}
({\bf k}_1,{\bf k}_2;{\bf k}_2-{\bf k}_1-{\bf q}) \nonumber
\end{eqnarray}
\end{mathletters}
So, small- and large-transfer scales  are intrinsically related due
to the Pauli principle, and any calculation of the low-energy corrections
should take this into account. Eventually the
integration over whole range of the transfer values through a
given loop results in the temperature corrections
not only to the vertex, but to the response functions as well,
when that loop is a part of a more complex diagram. The big transfer
($\sim 2k_F$) contributions essentially gave rise to the spin 
susceptibility  result (\ref{sus}) in the three-loop approximation.

We find that the relationships known from the classical FLT 
derivations at $T=0$ for the parameters of 
Galilean-invariant FL  (e.g, the effective mass, response functions 
vz components of the Landau function)  are violated by 
finite-temperature terms.
The coefficients in the temperature corrections
to these relationships subtly involve contrubutions from small and
large ($\sim 2k_F$) momentum processes.

Concerning the argument for the cancellation of anomalous 
terms in the response functions due to Ward identities:  
We have calculated the vertices at the one-loop level
${\cal O}(u^2)$. Through the Ward identities  the self-energy
corrections were taken into account with the same accuracy. There
are no more terms of the order ${\cal O}(u^2)$ to cancel the
temperature dependence (\ref{sus}). Thus, the linear temperature
dependence of susceptibility (or weaker temperature dependence of
the compressibility) does not contradict the {\it exact } Ward
identities known to us, moreover in our results for the response
functions both vertex and self-energy corrections are included on
the same footing {\it by using the Ward identities}.

For more realistic models of electrons in (quasi)-2D crystals, i.e., 
for various tight-binding spectra and fillings, 
the free-gas-like square-root $2k_F$ 
singularities (with $k_F$ depending on a chosen direction in ${\bf q}$-space) 
are known to exist in the Lindhard functions.\cite{Sher93}
We think this is enough to result in linear $T$-terms in physical 
quantities analogous to what we found in this study.
We argue  that the cancellation of the $T$-terms 
in some FL parameters is special to second order perturbation theory
and the model considered, while the leading linear temperature 
corrections are a generic feature of the 2D FL.

We hope our results may be experimentally tested in real 2D FL systems.
For example, a very naive fit of the temperature dependence of the spin 
susceptibility in ${\rm Sr_2RuO_4}$ system\cite{Maeno} when it is in the 
2D metallic regime (above 3D crossover temperature) shows that the data are  
compatible with the form (\ref{sus}). 
We expect our results stimulate a more detailed
examination of the leading temperature dependences
of response functions in 2D systems.
%
%
%%%%%%%%%%%%%%%%%%%%%%%%%%%%%%%%%%%%%%%%%%%%%%%%%%%%%%%%%%%%%%%%%%%%%%%%%%%%%%
%%%%%%%%%%%%%%%%%%%%%%%%%%%%%%%%%%%%%%%%%%%%%%%%%%%%%%%%%%%%%%%%%%%%%%%%%%%%%%
%------------------------------------------------------------------------------
\acknowledgements
Stimulating conversations with S. Lukyanov are gratefully acknowledged.
We thank V. Oudovenko for his help with the numerical tests of our results.
This work is supported by NSF DMR0081075.
%

%%%%%%%%%%%%%%%%%%%%%%%%%%%%%%%%%%%%%%%%%%%%%%%%%%%%%%%%%%%%%%%%%%%%%%%%%%%%%%
\begin {appendix}
\section{Zero sound (ZS) and Peierls (ZS') graphs }\label{ApA}
%%%%%%%%%%%%%%%%%%%%%%%%%%%%%%%%%%%%%%%%%%%%%%%%%%%%%%%%%%%%%%%%%%%%%%%%%%%%%%
%
%
\subsection{Finite-temperature Lindhard function}
We define the Lindhard function as
\begin{equation}
\label{Lin}
L({\cal Q}) \equiv L({\bf Q},\Omega_n)
={1\over \beta} \sum_{i\omega_m} \int {d{\bf K}\over
(2\pi)^2}~
G_0({\bf K},\omega_m) ~
G_0({\bf K}+{\bf Q},\omega_m+{\Omega_n })
\end{equation}
It corresponds to the contribution of the particle-hole loop
in graph ZS or ZS', depending on the actual value of
${\cal Q}$ we put.
{\it For the purpose of notational economy in the Appendices
we will reserve capital letters for momentum vectors, while
small letters correspond to their dimensionless counterparts
defined below.}
The free electron Green's function entering the loop is given by
\begin{equation}
\label{G0def}
G_0^{-1}({\bf K},\omega_n) = i\omega_n+\mu-\epsilon({\bf K})
\end{equation}
with its spectrum
\begin{equation}
\label{spectrum}
\epsilon({\bf K})={K^2 \over 2m},~~\mu={K_F^2 \over 2m},~~
\xi_{\bf K} \equiv \epsilon({\bf K})- \mu
\end{equation}
A straightforward summation over Matsubara frequencies results in
\begin{equation}
\label{Lin2}
L({\cal Q})=\int {d{\bf K}\over (2\pi)^2}~
{n_{\circ} \big( \xi_{\bf K} \big)-
 n_{\circ} \big( \xi_{\bf{K}+\bf{Q} } \big)
\over
i\Omega_n +\xi_{\bf K}-\xi_{\bf K+\bf{Q}  }  }
\end{equation}
We denote by $n_{\circ}$ the Fermi-Dirac distribution function
\begin{equation}
\label{FerTh}
n_{\circ}(x) \equiv {1 \over e^{\beta x}+1}=
\frac12 -\frac12 \tanh \Big( {\beta x \over 2} \Big)
\end{equation}
We choose 2D polar axis along the transfer vector such that
${\bf KQ}=KQ \cos\vartheta $.
Then the Lindhard function can be transformed to the expression
\begin{equation}
\label{L0shift}
L({\bf Q},0)=-{m \over \pi^2}
\int_0^{\infty}K dK n_{\circ} \big( \xi_{\bf K} \big)
 \int_0^{2 \pi}d \vartheta
{1 \over Q^2-4K^2 \cos^2\vartheta}
\end{equation}
The angular integration above is understood in the sense of the
principal value.
We define the dimensionless Lindhard function $\chi$ at zero
frequency transfer as
\begin{equation}
\label{Lchi}
L({\bf Q},0) \equiv -\nu_0 \chi({\bf Q}), ~~\nu_0={m \over 2\pi}
\end{equation}
where $\nu_0$ is the free 2D electron density of states per spin.
Let us also introduce the new (dimensionless) variables
\begin{equation}
\label{dmlsvar}
\beta_0=  {\beta K_F^2 \over 4m}, ~ q={ Q \over 2K_F},~
 k={ K \over K_F}
\end{equation}
In order to proceed we use the following result for the principal
value of the integral $I_a$ at $a>0$:
\begin{equation}
\label{Ia}
I_a  \equiv \int_0^{2 \pi}
{ d\vartheta \over a^2-\cos^2\vartheta }=
{2 \pi \over a \sqrt{a^2-1}}~ \Theta(a^2>1)
\end{equation}
where $\Theta$ is the event (Heaviside) function.
After the angular integration we obtain
\begin{equation}
\label{chi2s}
\chi(q)={1 \over 2  q}
\int_0^q {k dk \over \sqrt{q^2-k^2} }
\big\{ 1-\tanh \beta_0(k^2-1) \big\}
\end{equation}
The above formula can be easily casted into the final form
\begin{equation}
\label{chi8}
\chi(q)=\frac12+ \frac12  \tanh \beta_0
-{\beta_0 \over 2}
\int_{-1}^{q^2-1}{ dz \over\cosh^2 \beta_0 z}
\sqrt{1-{1+z \over q^2}}
\end{equation}
To make zero-temperature check of our result the following
formula is useful
\begin{equation}
\label{T0delta}
\beta_0 \to \infty:~~ \frac12{ \beta_0 \over\cosh^2 \beta_0 z}
\to \delta(z)
\end{equation}
Then from Eq.~(\ref{chi8}) we easily recover the result known for $T=0$,
which was first found by Stern:\cite{Stern}
\begin{mathletters}
\label{LT0}
\begin{eqnarray}
q^2&<&1:~~\chi(q)=1 \\
q^2&>&1:~~\chi(q)=1-\sqrt{1-{1 \over q^2}}
\end{eqnarray}
\end{mathletters}
%%
%%%%%%%%%%%%%%%%%%%%%%%%%%%%%%%%%%%%%%%%%%%%%%%%%%%%%%%%%%%%%%%%%%%%%%%%%%%%%%
\subsection{Fourier components}
Let us consider now the Lindhard function at the ``effective'' transfer
\begin{equation}
\label{Qprime}
{\bf Q'}={\bf K}_F^{(2)}-{\bf K}_F^{(1)}-{\bf Q}
\end{equation}
which occurs in the ZS' loop when the ``real'' transfer in the
vertex is ${\bf Q}$. Parameterizing vectors ${\bf K}_F^{(2)}$ and
${\bf K}_F^{(1)}$ by their angles ($\theta_1$, $\theta_2$)
relatively to the direction of vector ${\bf Q}$, we get
\begin{equation}
\label{Qprime2}
{ Q'}^2={ Q}^2 +4K_F^2\sin^2{\theta_2-\theta_1 \over 2}
+4QK_F\sin{\theta_2-\theta_1 \over 2}\sin{\theta_2+\theta_1 \over 2}
\end{equation}
In terms of new angles
\begin{equation}
\label{angles}
\theta \equiv {\theta_2-\theta_1 \over 2}, ~~
\psi \equiv {\theta_2+\theta_1 \over 2}
\end{equation}
and of the dimensionless variables (\ref{dmlsvar})
one reads
\begin{equation}
\label{QprimeDls}
{ q'}^2={ q}^2+ \sin^2\theta +2 q \sin\theta \sin\psi
\end{equation}
Let us consider now $\chi(q')$ at zero vertex transfer $q=0$, i.e.,
${ q'}^2=q_{12}^2=\sin^2\theta $. For the Fourier transform
\begin{equation}
\label{Fn}
{\chi}_n ={1 \over \pi}
\int_0^{\pi}d\theta \cos(2m \theta) \chi(\theta)
\end{equation}
of $\chi(\theta)$ given by Eq.(\ref{chi8}), we have
\begin{equation}
\label{chiFour}
\chi_n=
\biggl( \frac12+ \frac12  \tanh \beta_0 \biggr)
\delta_{n,0}-\delta\chi_n
\end{equation}
where
\begin{equation}
\label{deltachiN}
\delta\chi_n \equiv
{ \beta_0 \over 2 \pi}
\int_0^{\pi}d\theta \cos(2n \theta)
\int_{-1}^{\sin^2\theta-1}{ dz \over\cosh^2 \beta_0 z}
\sqrt{1-{1+z \over \sin^2\theta}}
\end{equation}
Then
\begin{equation}
\label{dchi0}
\delta\chi_0 =
{ \beta_0 \over \pi}
\int_0^1{dx \over x \sqrt{1-x^2}}
\int_{-1}^{x^2-1}{ dz \over\cosh^2 \beta_0 z}
\sqrt{x^2- (1+z) }=
{ \beta_0 \over \pi}
\int_{-1}^0{ dz \over\cosh^2 \beta_0 z}
\int_{\sqrt{1+z}}^1
{dx \over x \sqrt{1-x^2}}
\sqrt{x^2- (1+z) }
\end{equation}
Using the result
\begin{equation}
\label{useint}
\int_{a}^1
{dx \over x \sqrt{1-x^2}}
\sqrt{x^2-a^2}= {\pi \over 2}(1-a)
\end{equation}
we get
\begin{equation}
\label{dchi01}
\delta\chi_0 =
 \frac12  \tanh \beta_0 -
{ \beta_0 \over 2}
\int_{-1}^0{ dz \over\cosh^2 \beta_0 z} \sqrt{1+z}
\end{equation}
Thus
\begin{equation}
\label{chi0ex}
\chi_0=\frac12+
{ \beta_0 \over 2}
\int_{0}^1{ dz \over\cosh^2 \beta_0 z} \sqrt{1-z}=
\frac12+
\frac12\int_{0}^{1 \over T_0}{ dz \over\cosh^2 z}
\sqrt{1-T_0 z}
\end{equation}
The dimensionless temperature $T_0$ is
\begin{equation}
\label{T0}
T_0 \equiv {1 \over \beta_0} \equiv {4m T \over K_F^2} \ll 1
\end{equation}
Taylor-expanding the square root under the integral, we can
then extend the upper limit to the infinity with exponential
accuracy. Thus,
\begin{equation}
\label{chi0Tay}
\chi_0 \approx \frac12+
\frac12 \int_{0}^{\infty}{ dz \over\cosh^2 z}
\biggl( 1-\frac12 T_0z -\frac18 T_0^2 z^2 -{1 \over 16} T_0^3 z^3
+{\cal O}(T_0^4) \biggr)
\end{equation}
After simple integration we arrive to the sought result:
\begin{equation}
\label{chi0app}
\chi_0 = 1-{\ln 2 \over 4} T_0- {\pi^2 \over 192}T_0^2
-{9 \zeta(3) \over 256} T_0^3+{\cal O}(T_0^4)
\end{equation}
For the next Fourier component using $\cos2\theta=1-2\sin^2\theta$
we obtain
\begin{equation}
\label{chi1}
\chi_1=-\delta\chi_0 +
{2 \beta_0 \over \pi}
\int_{-1}^0{ dz \over\cosh^2 \beta_0 z}
\int_{\sqrt{1+z}}^1
{xdx \over  \sqrt{1-x^2}}
\sqrt{x^2- (1+z) }
=-\delta\chi_0 + { \beta_0 \over 2}
\int_{0}^1{ z dz \over\cosh^2 \beta_0 z}
\end{equation}
Thus,
\begin{equation}
\label{chi1b}
\chi_1 = -\delta\chi_0+{\ln 2 \over 2} T_0 +{\cal O}({\rm e}^{-1/T_0})
\end{equation}
and combining the last formula with the previous results, we have
\begin{equation}
\label{chi1app}
\chi_1 = {\ln 2 \over 4} T_0- {\pi^2 \over 192}T_0^2
-{9 \zeta(3) \over 256} T_0^3+{\cal O}(T_0^4)
\end{equation}
%
%
%%%%%%%%%%%%%%%%%%%%%%%%%%%%%%%%%%%%%%%%%%%%%%%%%%%%%%%%%%%%%%%%%%%%%%%%%%%%%%
\section{BCS graph }\label{ApB}
%%%%%%%%%%%%%%%%%%%%%%%%%%%%%%%%%%%%%%%%%%%%%%%%%%%%%%%%%%%%%%%%%%%%%%%%%%%%%%
\subsection{BCS loop contribution at finite temperature}
We define the contribution of the particle-particle loop in the
BCS graph as
\begin{equation}
\label{Cooper}
{\Bbb   C}({\cal Q}_s) \equiv {\Bbb C}({\bf Q}_s,\Omega_s)
={1\over \beta} \sum_{i\omega_m} \int {d{\bf K}\over
(2\pi)^2}~
G_0({\bf K},\omega_m) ~
G_0({\bf- K}+{{\bf Q}_s } ,-\omega_m+{\Omega_s })
\end{equation}
where
\begin{equation}
\label{Qs}
{\bf Q}_s \equiv {\bf K}_1 +{\bf K}_2, ~~\Omega_s \equiv \omega_1+\omega_2
\end{equation}
After summation of the Matzubara frequencies we get
\begin{equation}
\label{Cooper2}
{\Bbb C}({\cal Q}_s)=-\int {d{\bf K}\over (2\pi)^2}~
{n_{\circ} \big(\xi_{{\bf K} }\big)
-n_{\circ} \big(-\xi_{{\bf- K}+{\bf Q}_s }\big)
\over
-i\Omega_s +\xi_{{\bf K}}+\xi_{{\bf- K}+{\bf Q}_s }  }
=\frac12 \int {d{\bf K}\over (2\pi)^2}~
{\tanh \big(\frac12 \beta\xi_{{\bf K}} \big)+
\tanh \big(\frac12 \beta \xi_{{\bf- K}+{\bf Q}_s } \big)
\over
-i\Omega_s + \xi_{{\bf K} }+\xi_{{\bf- K}+{\bf Q}_s  } }
\end{equation}
For the case we are interested in
 (i.e., zero sum frequency $\Omega_s$ and vectors ${\bf K}_1$, ${\bf K}_2$ lying on the
Fermi surface, cf. also  notations (\ref{Qprime}, \ref{angles})) we have
\begin{equation}
\label{QsFer}
 Q_s=|{\bf K}_F^{(2)}+{\bf K}_F^{(1)}|=2K_F \cos\theta, ~~\Omega_s=0
\end{equation}
To regularize  the momentum integral in (\ref{Cooper2}) we will introduce an
ultraviolet cutoff $\Lambda$ which we assume to be the largest momentum scale of
the problem. Under this condition  we shift the momentum integration in the second
term of the sum on the r.h.s. of Eq.~(\ref{Cooper2}), obtaining the expression
\begin{equation}
\label{CoopSh}
{\Bbb C}({\cal Q}_s)= \frac12 \int {d{\bf K}\over (2\pi)^2}~
\Bigg(
{ \tanh \big(\frac12 \beta\xi_{{\bf K}} \big) \over
-i\Omega_s + \xi_{{\bf K} }+\xi_{{\bf- K}+{\bf Q}_s  } }
+
{ \tanh \big(\frac12 \beta\xi_{-{\bf K}} \big) \over
-i\Omega_s + \xi_{{\bf K}+{\bf Q}_s  }+\xi_{{\bf- K} } }
\Bigg)
\end{equation}
Two-dimensional integration in (\ref{CoopSh}) reads as:
\begin{equation}
\label{2Din}
 \int d{\bf K}=\int_0^{\Lambda} K dK
\int_0^{2\pi}d \vartheta
\end{equation}
Using the previously defined dimensionless parameters [cf. notations (\ref{dmlsvar})]
along with  new variables
\begin{equation}
\label{qsdmls}
q_s \equiv {Q_s \over 2K_F}~,~\lambda \equiv {\Lambda \over K_F} \gg 1
\end{equation}
and the dimensionless function $C$ [cf. notations (\ref{Lchi})]
\begin{equation}
\label{CC}
{\Bbb C} ({\bf Q}_s,0) \equiv \nu_0 C({\bf q}_s)
\end{equation}
one gets
\begin{equation}
\label{Cdls}
C(q_s)={1 \over 4 \pi }
\int_0^{\lambda}2 k dk \tanh \beta_0(k^2-1)
\int_0^{2 \pi} d\vartheta
{k^2-1+2q_s^2 \over
(k^2-1+2q_s^2)^2-4k^2q_s^2 \cos^2\vartheta }
\end{equation}
or
\begin{equation}
\label{Cs2}
C(q_s)={1 \over 4 \pi }
\int_{-1}^{\lambda^2-1}dz \tanh \beta_0 z~
{z+2q_s^2 \over 4(z+1)q_s^2}~ I_a
\end{equation}
We denote $I_a$ according to Eq.(\ref{Ia}) with
\begin{equation}
\label{a}
a^2 \equiv
{(z+2q_s^2)^2 \over 4(z+1)q_s^2} \geq 0
\end{equation}
Note that
\begin{equation}
\label{avsz}
a^2>1 ~ \Longleftrightarrow z^2>4q_s^2(1-q_s^2)
\end{equation}
After some simple manipulations we get
\begin{equation}
\label{Cs3}
C(q_s)=\frac12 \int_{-1}^{\lambda^2-1}dz
{\tanh \beta_0 z \over \sqrt{ z^2-4q_s^2(1-q_s^2)} }
~{\rm sign}(z+2q_s^2)~\Theta(z^2>4q_s^2(1-q_s^2) )
\end{equation}
The square root is chosen to be positive on the real axis of integration $z$.
Recalling that $q_s^2=\cos^2\theta$ and, consequently,
\begin{equation}
\label{qSin}
4q_s^2(1-q_s^2)=\sin^2 2\theta~,
\end{equation}
we see that the sign- and event functions
play their role on the patch $z \in [-1,1]$ only. Due to the symmetry properties
of function $C(q_s=\cos\theta)$ it suffices to consider the angle range
$0 \leq \theta \leq \pi/2$.  After some analysis one can obtain
\begin{equation}
\label{CqF}
C(q_s)=
 \frac12 \int_{1}^{\lambda^2-1}dz
{\tanh \beta_0 z \over \sqrt{ z^2-\sin^2 2\theta} }
+
\Theta \Big(
{\pi \over 4} < \theta \leq {\pi \over 2}
\Big)
\int_{\sin 2\theta}^{1}dz
{\tanh \beta_0 z \over \sqrt{ z^2-\sin^2 2\theta} }~
\end{equation}
The function $C(q_s)$ is continuous. It can be easily calculated at
zero temperature, giving the known result (See. e.g.,
Ref.~[\onlinecite{JAPS}]):
\begin{equation}
\label{CqT0}
C(q_s)= \frac12 \ln {\lambda^2 \over q_s^2}~,~~ 0<q_s<1~,
\end{equation}
as well as for zero total incoming momentum ($q_s=0$) at low ($T_0
\ll 1$) temperature, giving with exponential precision
\begin{equation}
\label{Cq0}
C(0)=  \ln \Big(
 {4{\rm e}^{\gamma} \over \pi} {\lambda \over T_0}
\Big)
\end{equation}
where $\gamma =0.577..$ is Euler's constant.
For the latter calculation cf., e.g., section 33.3 of
Ref.~[\onlinecite{AGD}]. In the above formulas and in what follows
we track $\lambda$ up to terms ${\cal O}(1)$.
%%%%%%%%%%%%%%%%%%%%%%%%%%%%%%%%%%%%%%%%%%%%%%%%%%%%%%%%%%%%%%%%%%%%%
%%%%%%%%%%%%%%%%%%%%%%%%%%%%%%%%%%%%%%%%%%%%%%%%%%%%%%%%%%%%%%%%%%%%%
\subsection{Fourier components}
From Eq.~(\ref{CqF}) it is easy to calculate Fourier components of
function $C(q_s)$ given by
\begin{equation}
\label{CuF}
C_n = {2\over \pi} \int_0^{\pi \over 2}
d\theta ~\cos2n\theta~C(\theta)
\end{equation}
For the zeroth component after some manipulations we have
\begin{equation}
\label{CuF0}
C_0= {1 \over \pi}
\int_0^1 dz \tanh \beta_0 z
\int_0^{\arcsin z} {d \psi \over \sqrt{ z^2-\sin^2 \psi} }
+
{1 \over \pi}
\int_0^{\pi \over 2} d \psi
\int_{1}^{\lambda^2-1}dz
{\tanh \beta_0 z \over \sqrt{ z^2-\sin^2 \psi} }
\equiv C_0^{(1)}+C_0^{(2)}
\end{equation}
The first term can be written as
\begin{equation}
\label{C01}
 C_0^{(1)}={1 \over \pi}
\int_0^1 dz \tanh \beta_0 z ~K(z)
\end{equation}
where $K(z)$ is the complete elliptic integral of the first kind, which
has the following Taylor expansion:
\begin{equation}
\label{EllTaylor}
K(z)={\pi \over 2} \Big(
1+ \frac14 z^2 +{9 \over 64}z^4 + {\cal O}(z^6) \Big)
\end{equation}
To proceed further we notice that
\begin{equation}
\label{EllInt}
\int K(z)={\pi \over 2} z \cdot
{~}_3F_2 \Big[ \Big\{ \frac12,\frac12,\frac12 \Big\},
\Big\{ 1, \frac32 \Big\}; z^2 \Big]
\equiv {\Bbb F}(z)
\end{equation}
where the Taylor expansion of the hypergeometric function can be readily
read from  the term by term integration of the expansion (\ref{EllTaylor}),
i.e.,
\begin{equation}
\label{HTaylor}
{\Bbb F}(z) ={\pi \over 2}  \Big(
z +{1 \over 12}z^3 +{9 \over 320}z^5+ {\cal O}(z^7)
\Big)
\end{equation}
Integrating by parts the r.h.s. of Eq.~(\ref{C01}) we obtain
\begin{equation}
\label{C01bp}
 C_0^{(1)}={1 \over \pi} \Big(
{\Bbb F}(1) \cdot \tanh \beta_0-
\int_0^1 {\beta_0 dz \over \cosh^2\beta_0z} {\Bbb F}(z)
\Big)
\end{equation}
Using the same (exponentially accurate at low temperature) approximations
which lead us to Eq.~(\ref{chi0Tay}), we end up
\begin{equation}
\label{C01bpTay}
C_0^{(1)} \approx {1 \over \pi}{\Bbb F}(1)
- \frac12 \int_0^{\infty} {dz \over\cosh^2z}
\Big( T_0z+{T_0^3 \over 12}z^3+{9 T_0^5 \over 320}z^5+ {\cal O}(T_0^7)  \Big)
\end{equation}
In order to calculate ${\Bbb F}(1)$ it is convenient to come back to the
integral representation of the elliptic integral, and then to evaluate it
via changing the order of integration.
\begin{equation}
\label{Hyp1}
{\Bbb F}(1)=
%\int_0^1 dz  K(z)=
\int_0^1 dz
\int_0^z {dx \over \sqrt{1-x^2}}{1 \over\sqrt{z^2-x^2} }
=\int_0^1 {dx \over \sqrt{1-x^2}}
\int_x^1 {dz \over\sqrt{z^2-x^2} }=
2 \int_0^{\pi \over4} d \theta
\ln \cot \theta =2 {\Bbb G}
\end{equation}
Here ${\Bbb G}=0.91569...$ is Catalan's constant. Finally
\begin{equation}
\label{C01F}
C_0^{(1)}=
{2 \over \pi}{\Bbb G} -{\ln2 \over 2}T_0 -
{3 \zeta(3) \over 64}T_0^3-{\cal O}(T_0^5)
\end{equation}
For the second contribution $C_0^{(2)}$ one integration by parts gives
\begin{equation}
\label{C02bp}
C_0^{(2)}={1 \over \pi}
\int_0^{\pi \over2}  d \psi
\Big(
\ln {2\lambda^2 \over 1+\cos \psi} -
\int_1^{\lambda^2-1}
{\beta_0 dz \ln(z+\sqrt{z^2-\sin^2 \psi}) \over \cosh^2 \beta_0 z}
\Big)
\end{equation}
It is easy to see that the second term on the r.h.s of Eq.~(\ref{C02bp})
is exponentially small, while the first one can be readily evaluated
resulting in
\begin{equation}
\label{C02F}
C_0^{(2)}= \ln2\lambda -{2 \over \pi}{\Bbb G}
\end{equation}
Combining our results together, we have
\begin{equation}
\label{C0F}
C_0=\ln2\lambda -{\ln2 \over 2}T_0 -
{3 \zeta(3) \over 64}T_0^3-{\cal O}(T_0^5)
\end{equation}
Calculation of  the first Fourier component $C_1$ is very simple, since
the first term on the r.h.s. of Eq.~(\ref{CqF}) gives zero after the
angular integration, while the second one can be evaluated exactly,
resulting in
\begin{equation}
\label{C1F}
C_1=-{1 \over 2 \beta_0} \ln \cosh \beta_0
= -\frac12 +{\ln2 \over 2}T_0 +{\cal O}({\rm e}^{-1/T_0})
\end{equation}
\end{appendix}
%%%%%%%%%%%%%%%%%%%%%%%%%%%%%%%%%%%%%%%%%%%%%%%%%%%%%%%%%%%%%%%%%%%%%%%%%%%%%%
%%%%%%%%%%%%%%%%%%%%%%%%%%%%%%%%%%%%%%%%%%%%%%%%%%%%%%%%%%%%%%%%%%%%%%%%%%%%%%

%%%%%%%%%%%%%%%%%%%%%%%%%%%%%%%%%%%%%%%%%%%%%%%%%%%%%%%%%%%%%%%%%%%%%%%%%%%%%%
% REFERENCES
%

%xxxxxxxxxxxxxxxxxxxxxxxxxxxxxxxxxxxxxxxxxxxxxxxxxxxxxxxxxxxxxxxxxxxxxxxxxxxxxx
\begin{figure}
\vglue 0.4cm\epsfxsize 10cm\centerline{\epsfbox{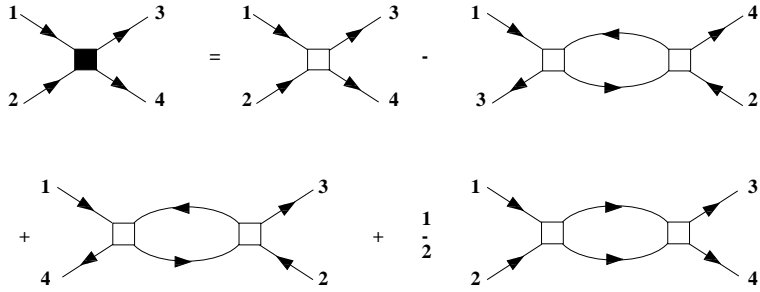}}\vglue 0.4cm
\caption{
Diagrammatic equation for  the four-point vertex at one-loop level.
The one-loop graphs are called ZS, ZS', and BCS in the order they appear
on the r.h.s. of this equation.}
\end{figure}
%xxxxxxxxxxxxxxxxxxxxxxxxxxxxxxxxxxxxxxxxxxxxxxxxxxxxxxxxxxxxxxxxxxxxxxxxxxxxxx

%%%%%%%%%%%%%%%%%%%%%%%%%%%%%%%%%%%%%%%%%%%%%%%%%%%%%%%%%%%%%%%%%%%%%%%%%%%%%%
%%%%%%%%%%%%%%%%%%%%%%%%%%%%%%%%%%%%%%%%%%%%%%%%%%%%%%%%%%%%%%%%%%%%%%%%%%%%%%
%%%%%%%%%%%%%%%%%%%%%%%%%%%%%%%%%%%%%%%%%%%%%%%%%%%%%%%%%%%%%%%%%%%%%%%%%%%%%%
%%%%%%%%%%%%%%%%%%%%%%%%%%%%%%%%%%%%%%%%%%%%%%%%%%%%%%%%%%%%%%%%%%%%%%%%%%%%%%
%%%%%%%%%%%%%%%%%%%%%%%%%%%%%%%%%%%%%%%%%%%%%%%%%%%%%%%%%%%%%%%%%%%%%%%%%%%%%%
%%%%%%%%%%%%%%%%%%%%%%%%%%%%%%%%%%%%%%%%%%%%%%%%%%%%%%%%%%%%%%%%%%%%%%%%%%%%%%
%%%%%%%%%%%%%%%%%%%%%%%%%%%%%%%%%%%%%%%%%%%%%%%%%%%%%%%%%%%%%%%%%%%%%%%%%%%%%%
%%%%%%%%%%%%%%%%%%%%%%%%%%%%%%%%%%%%%%%%%%%%%%%%%%%%%%%%%%%%%%%%%%%%%%%%%%%%%%
%%%%%%%%%%%%%%%%%%%%%%%%%%%%%%%%%%%%%%%%%%%%%%%%%%%%%%%%%%%%%%%%%%%%%%%%%%%%%%
%==============================================================================

\begin{references}
\bibitem{note3} We will not present details of these
calculations, since they are too technical. The reader may work them out
along the lines of the simpler analytical procedure described in the
Appendices for the one-loop vertex corrections.
\end{references}
\end{document}